\documentclass[%
 aps,
 amsmath,amssymb,
 reprint,%
]{revtex4-2}

\usepackage{graphicx}
\usepackage{dcolumn}
\usepackage{bm}

\usepackage{mathptmx}
\usepackage{comment}
\usepackage{etoolbox}
\usepackage{bbm}
\usepackage{xcolor}
\usepackage{tikz}
\usepackage[utf8]{inputenc} 
\usepackage[T1]{fontenc}    
\usepackage{hyperref}       
\usepackage{url}            
\usepackage{booktabs}       
\usepackage{amsfonts}       
\usepackage{nicefrac}       
\usepackage{microtype}      
\usepackage{lipsum}
\usepackage{fancyhdr}       
\usepackage{algorithmic}
\usepackage{algorithm}
\graphicspath{{media/}}     

\usepackage{CJKutf8}

\draft 

\begin{document}



\preprint{AIP/123-QED}

\title{Nonlinear Bayesian Doppler Tomography for Simultaneous Reconstruction of Flow and Temperature
}



\author{Kenji Ueda}
\email{ueda.kenji@nifs.ac.jp}
\affiliation{National Institute of Fusion Science}

\author{Masaki Nishiura}
\affiliation{National Institute of Fusion Science}
\affiliation{The University of Tokyo}


\date{\today}

\begin{abstract}
We present a nonlinear Bayesian tomographic framework for Doppler spectral imaging that enables simultaneous reconstruction of emissivity, ion temperature, and flow velocity from line-integrated spectra. The method employs nonlinear Gaussian process tomography (GPT) with a Laplace approximation while retaining the full Doppler forward model.
A log-Gaussian process prior stabilizes the velocity reconstruction in low-emissivity regions where Doppler information becomes weak, preventing the unphysical divergence of velocity estimates commonly encountered in conventional spectral tomography. The reconstruction method is verified using synthetic phantom data and applied to coherence imaging spectroscopy (CIS) measurements in the RT-1 device, resolving spatial structures of ion temperature and toroidal ion flow characteristic of magnetospheric plasma in the RT-1 device. The framework extends existing CIS tomography to regimes with strong flows and large temperature variations and provides a general Bayesian approach for Doppler spectral tomography that can be integrated with complementary spectroscopic diagnostics.
\end{abstract}

\maketitle
\section{Introduction}\label{sec:Introduction}

Doppler spectral imaging is a widely used technique for inferring local temperature and flow velocity in spatially distributed media. Such measurements are central to a broad range of physical systems, including Doppler tomography of accretion disks in astrophysics \cite{Marsh1988}, atmospheric wind measurements using Doppler lidar and radar \cite{Rye1989}, biomedical blood-flow diagnostics based on Doppler ultrasound imaging \cite{Jensen1996}, and spectroscopic measurements of ion velocity and temperature in laboratory plasmas \cite{Fonck1984,Howard_2003}.
In all these applications, physical quantities are encoded in spectral shifts and line broadening arising from Doppler effects.

In many practical situations, however, the measured signals do not directly represent local quantities. Instead, the observations correspond to line-integrated projections of spatially distributed variables along the measurement path. 
Recovering local temperature and velocity fields from such data therefore constitutes a nonlinear inverse problem. 
The forward model typically couples emissivity, temperature-dependent Doppler broadening, and velocity-induced spectral shifts, leading to strong parameter entanglement and ill-posedness of the reconstruction.

Tomographic reconstruction techniques have long been used to address such inverse problems by exploiting multiple lines of sight and prior assumptions on spatial structure \cite{Tarantola2005}. In particular, Bayesian inference provides a systematic framework to combine prior knowledge with measurement likelihoods, allowing statistically consistent estimation of spatial distributions together with uncertainty quantification.

Gaussian process (GP) models offer a flexible and nonparametric representation of spatial correlations and have been successfully applied to tomographic inversion problems \cite{GP_for_ML}. In the context of plasma diagnostics, Gaussian Process Tomography (GPT) has been introduced as a powerful approach for reconstructing emissivity distributions while naturally incorporating spatial smoothness and measurement uncertainty \cite{J_Svensson_2011, Li2021-ce}. However, most existing applications rely on linearized forward models or assume small Doppler shifts\cite{Howard2010-yt,Silburn_MAST_2014,Meyer_2018,Bingli_Li_2022_CIS_tomography}, which limits their applicability in regimes with strong flows or large temperature variations.

In this work, we develop a nonlinear Bayesian tomographic framework that directly incorporates the full Doppler forward model without linearization. The method employs nonlinear Gaussian Process Tomography~combined with the Laplace approximation to estimate the joint posterior distribution of emissivity, ion temperature, and flow velocity. By modeling these quantities as correlated Gaussian processes and analytically marginalizing latent variables, the framework enables simultaneous reconstruction of multiple physical quantities with rigorous uncertainty quantification.

Although the methodology is general and applicable to a wide range of Doppler-based imaging diagnostics, we demonstrate it here using coherence imaging spectroscopy (CIS) measurements in the RT-1 laboratory plasma device~ \cite{Nishiura_2019}. 
The proposed framework enables stable reconstruction in regimes with strong Doppler shifts and temperature variations and resolves long-standing issues such as velocity divergence in low-emissivity regions.

This paper is organized as follows. Section~\ref{sec:GPT} introduces the Bayesian tomographic framework and Gaussian-process prior. Section~\ref{sec:CIS} describes the CIS measurement principle and derives the nonlinear projection equations. Section~\ref{sec:Nonlinear_GPT_CIS} develops the nonlinear Bayesian model used for inference. Section~\ref{sec:Test_Phantom_Data} validates the method using synthetic phantom data, and Section~\ref{sec:Tomography_Experimental_Data} demonstrates the reconstruction using experimental CIS measurements in the RT-1 device.

The present work therefore establishes a general Bayesian inversion framework for Doppler spectral imaging that is applicable not only to plasma diagnostics but also to a wide range of Doppler-based measurement systems in applied physics, including astrophysical observations, atmospheric remote sensing, and biomedical flow imaging.

\section{Gaussian Process Tomography}\label{sec:GPT}
GPT provides a probabilistic framework for solving inverse problems with spatially correlated priors \cite{GP_for_ML, Calvetti2007, Tarantola2005}. In this framework, the spatial distributions of the physical quantities are modeled as Gaussian processes, allowing regularization of the ill-posed tomographic inversion while preserving spatial correlations.
This section summarizes the Bayesian framework and Gaussian-process prior used for tomography. We then describe the CIS measurement principle and derive the nonlinear projection equations in Section~\ref{sec:CIS}. These elements are combined into the CIS-specific nonlinear GPT model and inference procedure in Section~\ref{sec:Nonlinear_GPT_CIS}.

\subsection{Bayesian Tomography Using Gaussian Processes} 
In plasma diagnostics, one of the primary objectives is to reconstruct an unknown local quantity \( f(\vec{r}) \), such as emissivity, temperature, or velocity, from observed data \( d(\vec{x}) \), where \( \vec{r} \) represents positions within the plasma, and \( \vec{x} \) denotes sensor coordinates. 
Bayesian tomography provides a systematic framework for this inverse problem by incorporating prior knowledge and observed data to estimate the posterior probability of the unknown quantity.
According to Bayes' theorem, the posterior probability of \( f \) given the data \( d \) and hyperparameters \( \theta \) is expressed as:
\begin{eqnarray} 
\overbrace{{p(f \mid d,\theta)}}^{\mathrm{posterior}}
&=&\frac{\overbrace{p(d\mid f ,\theta )}^{\mathrm{likelihood}}\times\overbrace{p(f\mid \theta)}^{\mathrm{prior}}}{\underbrace{p(d\mid \theta )}_{\mathrm{evidence}}},\nonumber\\
&\propto& p(d\mid f,\theta)\times p(f\mid \theta).\label{eq: Bayes_theorem}
\end{eqnarray}

In GPT~\cite{J_Svensson_2011,Dong_Li_2013,Wang_T_2018}, the prior probability \( p(f\mid \theta) \) is modeled as a Gaussian process. 
This implies that any finite collection of function values \( \bm{f} = \{f(\vec{r}_i)\}_{i=1}^{N} \) follows a multivariate normal distribution characterized by a mean vector \( \bm{\mu}_f \) and a covariance matrix \( K_f \), i. e., \(\bm{f} \sim \mathcal{N}(\bm{\mu}_{f}, K_f)\). 
The matrix \( K_f \) is constructed using a kernel function \( k\), with entities \(\{K_f\}_{i,j} = k(\vec{r}_i,\vec{r}_j)\).

The choice of kernel function \( k(\cdot,\cdot) \) is crucial, as it encodes our assumptions about the smoothness and spatial correlations of the unknown function \( f(\vec{r}) \). 
In this paper, we employ the Gibbs kernel~\cite{Gibbs_1997,Dong_Li_2013}, a generalization of the squared exponential (SE) kernel, defined as:
\begin{eqnarray}\label{eq: Gibbs_kernel}
  k{(\vec{r}_i, \vec{r}_j) }=
   \sigma_{f}^2  \left| \Sigma_{\ell}(\vec{r}_i) \right|^{\frac{1}{4}} \left| \Sigma_{\ell}(\vec{r}_j) \right|^{\frac{1}{4}} \left| \frac{\Sigma_{\ell}(\vec{r}_i) + \Sigma_{\ell}(\vec{r}_j)}{2} \right|^{-\frac{1}{2}} \nonumber\\
  \times \exp{\left( -\frac{1}{2} (\vec{r}_i - \vec{r}_j)^\mathrm{T} \left( \frac{\Sigma_{\ell}(\vec{r}_i) + \Sigma_{\ell}(\vec{r}_j)}{2} \right)^{-1} (\vec{r}_i - \vec{r}_j) \right)},
\end{eqnarray}
where \( \sigma_f^2 \) represents the signal variance, and \( \Sigma_{\ell}(\vec{r}) \) defines the local correlation length scale at location \( \vec{r} \), which is related to the inverse matrix of the metric tensor.
When \( \Sigma_\ell(\vec{r}) \) is a scalar multiple of the identity matrix, \( \Sigma_\ell(\vec{r}) = \ell^2 I \), the Gibbs kernel reduces to the standard SE kernel commonly used in Gaussian processes:
\begin{eqnarray*}
    k(\vec{r}_i, \vec{r}_j) = \sigma_f^2 \exp{\left( -\frac{|\vec{r}_i - \vec{r}_j|^2}{2 \ell^2} \right)}.
\end{eqnarray*}
However, in this study, we consider isotropic but non-uniform kernels to accommodate spatially varying correlation lengths within the plasma. Accordingly, the scaling matrix \( \Sigma_{\ell}(\vec{r}) \) is defined as:
\begin{eqnarray}
    \Sigma_{\ell}(\vec{r}) = \ell^2(\vec{r}) I = \begin{bmatrix}
        \ell^2(\vec{r}) & 0 \\
        0 & \ell^2(\vec{r}) 
    \end{bmatrix},
\end{eqnarray}
where \( \ell(\vec{r}) \) is the position-dependent length scale function, and \( I \) is the identity matrix. 
This formulation allows the kernel to adapt to local variations in the plasma, providing more flexibility in modeling spatial correlations.

\subsection{Nonlinear Gaussian Process Tomography}

In practical applications, the relationship between the observed data \( \bm{d} \) and the unknown local quantity \( \bm{f} \) is not always linear. 
Therefore, we consider the following general measurement model:
\begin{eqnarray}
  \bm{d} = \bm{g} (\bm{f}) +\bm{\epsilon},
\end{eqnarray}
where \( \bm{g}(\bm{f}) \) is a nonlinear function mapping the local quantity \( \bm{f} = \{f(\vec{r}_i)\}_{i=1}^N \) to the observed data \( \bm{d} = \{d(\vec{x}_i)\}_{i=1}^M \), and \( \bm{\epsilon} \) represents the measurement error, including random noise and systematic errors, assumed to be normally distributed with zero mean and covariance \( \Sigma_{g} \).

Assuming the prior probability of \( \bm{f} \) is Gaussian with mean \( \bm{\mu}_{f}^\mathrm{pri} \) and covariance \( K_{f} \), and the noise \( \bm{\epsilon} \) follows a Gaussian distribution with covariance \(\Sigma_g\), the posterior probability of \( \bm{f} \) given the data \( \bm{d} \) can be expressed as:
\begin{eqnarray} \label{eq: logP_log-GPT}
    \underbrace{\log{p(\bm{f}\vert \bm{d}, \theta)}}_{\mathrm{log\text{-}posterior}}
    &=&\underbrace{-\frac{1}{2}(  \bm{d} - \bm{g}(\bm{f}))^\mathrm{T} 
    \Sigma_{g}^{-1}
      (\bm{d} - \bm{g}(\bm{f}))}_{\mathrm{log\text{-}likelihood}}\nonumber\\
    &&\underbrace{-\frac{1}{2}(\bm{f}-\bm{\mu}_{f}^\mathrm{pri})^\mathrm{T} 
    K_{f}^{-1} (\bm{f}-\bm{\mu}_{f}^\mathrm{pri})}_{\mathrm{log\text{-}prior}} + \,C,
\end{eqnarray}
where $\log$ is natural logarithm, and \( C \) is a constant independent of \( \bm{f} \). 
The first term represents the log-likelihood, and the second term is the log-prior.

In the case where \( \bm{g}(\cdot) \) is a linear operator, the posterior distribution remains Gaussian, and analytical solutions are available, as seen in the standard GPT~\cite{J_Svensson_2011,Dong_Li_2013,Wang_T_2018}.
However, when \( \bm{g}(\cdot) \) is nonlinear, as in many practical situations, the posterior probability becomes non-Gaussian, and obtaining an analytical solution is intractable because Eq.~\eqref{eq: logP_log-GPT} is no longer a quadratic form in \( \bm{f} \).

To address this challenge, we employ the Laplace approximation~\cite{GP_for_ML,Kuss2005-va,Bishop2006}, which approximates the posterior probability by a Gaussian centered at the mode of the true posterior. 
Specifically, we denote:
\begin{eqnarray} 
    p(\bm{f}\mid \bm{d}, \theta) \overset{\mathrm{LA}}{\simeq} \mathcal{N}(\bm{f}\mid \tilde{\bm{\mu}}^\mathrm{LA},\tilde{\Sigma}^\mathrm{LA}),
\end{eqnarray}
where \( \tilde{\bm{\mu}}^\mathrm{LA} \) is the mode of the posterior distribution, and \( \tilde{\Sigma}^\mathrm{LA} \) is the inverse of the negative Hessian (second derivative) of the log-posterior evaluated at the mode. Mathematically, these are defined as:
\begin{eqnarray}
    \tilde{\bm{\mu}}^\mathrm{LA} &=& \underset{\bm{f}}{\arg\max} \ \Psi(\bm{f}),\label{eq: def_of_LA_mu}\\
    \tilde{\Sigma}^\mathrm{LA} &=& - \left[ \nabla^2 \Psi(\bm{f})\big|_{\bm{f}=\tilde{\bm{\mu}}^\mathrm{LA}} \right]^{-1},\label{eq: def_of_LA_Sigma}
\end{eqnarray}
where \(\Psi(\bm{f})\) is the unnormalized log-posterior function, given by \( \Psi(\bm{f}) \overset{\mathrm{const}}{=} \log{p(\bm{f}\mid \bm{d}, \theta)} \).
To find the mode \( \tilde{\bm{\mu}}^\mathrm{LA} \), we solve the optimization problem in Eq.~\eqref{eq: def_of_LA_mu}. 
This can be achieved using iterative methods such as the Newton-Raphson algorithm. 
The update rule for the Newton-Raphson method is given by:
\begin{eqnarray}\label{eq: Newtonmethod}
\tilde{\bm{f}}^{\,\mathrm{new}} = \tilde{\bm{f}}^{\,\mathrm{old}} 
- \alpha \left[ \nabla^2\Psi(\tilde{\bm{f}}^{\,\mathrm{old}}) \right]^{-1} \nabla \Psi(\tilde{\bm{f}}^{\,\mathrm{old}}),
\end{eqnarray}
where \( \tilde{\bm{f}}^{\,\mathrm{old}} \) is the current estimate, \( \tilde{\bm{f}}^{\,\mathrm{new}} \) is the updated estimate, \( \nabla \Psi(\tilde{\bm{f}}^{\,\mathrm{old}}) \) is the gradient of the log-posterior, \( \nabla^2\Psi(\tilde{\bm{f}}^{\,\mathrm{old}}) \) is the Hessian matrix, and \( \alpha \) is a step size parameter.
This iterative process is repeated until convergence, which is typically assessed when the norm of the gradient \( \|\nabla \Psi(\tilde{\bm{f}}^{\,\mathrm{new}})\| \) becomes sufficiently small, indicating that a local maximum has been found.
The step size \( \alpha \) can be set to 1 for simplicity, but choosing an optimal \( \alpha \) at each iteration can enhance convergence. 
Once the mode \( \tilde{\bm{\mu}}^\mathrm{LA} \) is obtained, the covariance \( \tilde{\Sigma}^\mathrm{LA} \) is computed using Eq.~\eqref{eq: def_of_LA_Sigma}. This provides an approximate Gaussian posterior distribution.

\section{Diagnostics with Coherence Imaging Spectroscopy}\label{sec:CIS}

\subsection{Measurement Principle}

This section summarizes the CIS measurement principle and derives the projection equations that link local emissivity, temperature, and velocity to the observed images. These equations provide the nonlinear forward model used in the subsequent tomography.

CIS is an imaging Doppler spectroscopy technique that converts spectral Doppler information into spatial fringe patterns using birefringent interferometry ~\cite{Howard_2003, Howard2010-fl}. The technique enables two-dimensional measurement of ion temperature and flow velocity with high spatial resolution and has been widely applied in laboratory plasma experiments~\cite{Howard2010-fl,Silburn_MAST_2014,Nakamura_2018, Perseo_Gradic_2020}. 
It produces fringe images by exploiting the interference patterns created due to phase differences introduced by birefringent crystals. 
The signal from the output image of CIS, \( S_\mathrm{CIS}(\vec{x}) \), is expressed~\cite{Howard_2003, Howard2010-fl}:
\begin{eqnarray}\label{eq: signals of CIS}
    S_\mathrm{CIS}(\vec{x}) =  I_0 + I_0 \zeta_\mathrm{I} \zeta_\mathrm{D} 
    \cos{ (\phi_{0} + \phi_{\mathrm{D}})}  + \epsilon.
\end{eqnarray}
It should be noted that each variable on the \emph{right-hand side} of Eq.~\eqref{eq: signals of CIS} is a function of \( \vec{x} \), but this dependence has been omitted for simplicity. 
In this equation, the first term, denoted as \( I_0 \), represents the bias component corresponding to the intensity of the incident light. 
The second term is the modulation component corresponding to the autocorrelation of the incident light and depends on \( I_0 \), \( \zeta_\mathrm{I} \), \( \zeta_\mathrm{D} \), \( \phi_{0} \), and \( \phi_{\mathrm{D}} \). 
Among these, the instrumental contrast and phase of the carrier fringe, \( \zeta_\mathrm{I} \) and \( \phi_{0} \), respectively, must be eliminated from the data by a suitable calibration technique.
The contrast factor, represented by the variable \( \zeta_\mathrm{D} \), is mainly affected by Doppler broadening of the spectrum, while the phase shift, represented by the variable \( \phi_{\mathrm{D}} \), is mainly affected by the Doppler shift. 

\subsection{Projection Equation}

\begin{figure}[htbp]
    \centering
    \includegraphics[width=85mm]{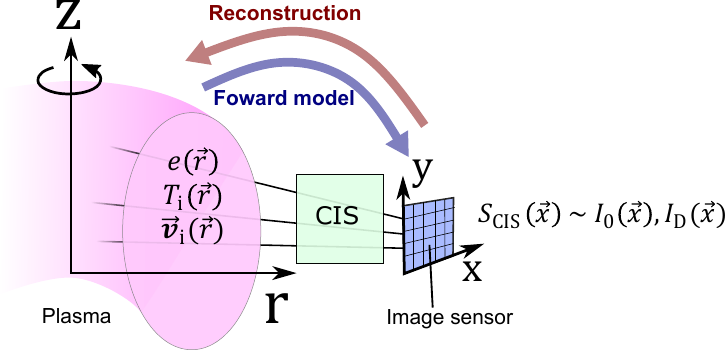}
    \caption{\label{fig: forward_model} 
    Conceptual diagram of Doppler spectral tomography using CIS. 
    Spatially distributed plasma parameters, including emissivity, ion temperature, and flow velocity, give rise to Doppler-broadened and Doppler-shifted spectral signals. 
    The CIS diagnostic records line-integrated projections of these spectra along multiple viewing chords. 
    The goal of the reconstruction is to infer the underlying spatial fields from these measurements within a Bayesian tomography framework.}
\end{figure}

The signal measured with CIS is the result of line integration of physical quantities including the emissivity \( e(\vec{r})\), ion temperature \( T_\mathrm{i} (\vec{r})\), and ion velocity \( \vec{v}_\mathrm{i}(\vec{r}) \) in the plasma, as shown in Fig.~\ref{fig: forward_model}.
Assuming a Maxwellian ion velocity distribution and radiation with a singlet spectral line, such as the 468~nm He~II line used in the present experiment, the CIS signal can be expressed as a line integral of the local emissivity, ion temperature, and ion velocity along the line of sight. Under these assumptions, the relation between the measured data, \( S_\mathrm{CIS}(\vec{x}) \), and the local plasma parameters \( T_\mathrm{i}(\vec{r}) \), \( \vec{v}_\mathrm{i}(\vec{r}) \), and \( e(\vec{r}) \) can be derived (see Appendix~\ref{app:derive_cis}):
\begin{eqnarray}
    I_{0}(\vec{x}) &=& \int_{L(\vec{x})} e(\vec{r}) \, \mathrm{d}l, \label{eq: transform_I0}\\
    I_\mathrm{D}(\vec{x}) 
    &=& \int_{L(\vec{x})} e(\vec{r}) \exp{\left[-\hat{T}(\vec{r})\right]}
    \exp{\left[ i \hat{\vec{v}}(\vec{r})\cdot \hat{\vec{l}}\right]} \, \mathrm{d}l, \label{eq: transform_TV}
\end{eqnarray}
where \( L(\vec{x}) \) denotes the line of sight (LOS) corresponding to the pixel at \( \vec{x} \), and \( \mathrm{d}l \) is the differential path length along \( L(\vec{x}) \). 
The left-hand sides correspond to the observed data acquired from Eq.~\eqref{eq: signals of CIS}, whereas the right-hand sides detail the line integrations performed along the LOS. 
\( I_\mathrm{D} \) is the effective amplitude of the modulated component of \( S_\mathrm{CIS} \), which is given by~\cite{Howard_2003, Howard2010-fl}:
\begin{eqnarray}
    I_\mathrm{D}(\vec{x}) = I_0(\vec{x}) \zeta_\mathrm{D}(\vec{x}) \exp{[i \phi_{\mathrm{D}}(\vec{x})]}.
\end{eqnarray}
In this paper, the hat symbol \( \hat{} \) denotes dimensionless quantities; \( \hat{T} \) and \( \hat{\vec{v}} \) are normalized as \( \hat{T} := T_\mathrm{i} / T_\mathrm{c} \) and \( \hat{\vec{v}} := \vec{v}_\mathrm{i} / v_\mathrm{c} \), respectively. Here, \( T_\mathrm{c} \) and \( v_\mathrm{c} \) are the characteristic temperature and characteristic velocity, given by~\cite{Howard_2003, Howard2010-fl}:
\begin{eqnarray}
    k_\mathrm{B} T_\mathrm{c} = 2 m_\mathrm{s} v_\mathrm{c}^2, \quad v_\mathrm{c} = \frac{c}{2\pi \hat{N}(\vec{x})},
\end{eqnarray}
where \( k_\mathrm{B} \) is the Boltzmann constant, \( c \) is the speed of light, \( m_\mathrm{s} \) is the mass of the ion species, and \( \hat{N}(\vec{x}) \) is group delay\cite{Gradic_2021}, defined as 
\begin{eqnarray}
    2\pi \hat{N} := -\lambda_{0} \left.\frac{d\phi}{d\lambda}\right|_{\lambda=\lambda_0},
\end{eqnarray}
where \( \lambda_0 \) is the central wavelength of the spectrum.
The first equation, Eq.~\eqref{eq: transform_I0}, is a standard projection equation, while the second equation, Eq.~\eqref{eq: transform_TV}, is a complex nonlinear equation where the emissivity, temperature, and velocity variables are intertwined, as discussed in the introduction. 

\subsection{Organizing Equations for the Tomography}

To simplify the solution of Eqs.~\eqref{eq: transform_I0} and~\eqref{eq: transform_TV}, we introduce new variables \( \hat{e} \) and \( \hat{a} \), defined as:
\begin{eqnarray}
    \hat{e} &:=& \log{e}, \label{eq: def_of_hate}\\
    \hat{a} &:=& \hat{e} - \hat{T}, \label{eq: def_of_a}
\end{eqnarray}
where we define \( \hat{e} \) and \( \hat{a} \) as the "log-emissivity" and the "local amplitude," respectively. 
By taking the logarithm of the emissivity, we combine the temperature and emissivity into a single variable, \( \hat{a} \). 
Furthermore, previous studies~\cite{Ueda2025-pc} have indicated that this approach improves the accuracy of tomography.

Thus, Eqs.~\eqref{eq: transform_I0} and \eqref{eq: transform_TV} can be rewritten and discretized for each \( i \) as follows:
\begin{eqnarray}
    \{\bm{g}_0(\hat{\bm{e}})\}_i    &:=& \sum_{j=1}^{N} H_{ij} \exp{\hat{e}_{j}},\label{eq: discretized_g0}\\
    \{\bm{g}_\mathrm{C}(\hat{\bm{a}},\hat{\bm{v}})\}_i &:=& \sum_{j=1}^{N}  H_{ij} \exp{\hat{a}_{j}} \cos[(\Theta_{ij} \hat{v}_j)],\label{eq: discretized_gc}\\
    \{\bm{g}_\mathrm{S}(\hat{\bm{a}},\hat{\bm{v}})\}_i &:=& \sum_{j=1}^{N}  H_{ij} \exp{\hat{a}_{j}} \sin[(\Theta_{ij} \hat{v}_j)],\label{eq: discretized_gs}
\end{eqnarray}
where \( H \) is the geometry matrix with dimensions \( M \times N \), \( M \) is the number of measurement points (pixels), \( N \) is the number of discretized points in the plasma domain, and \( \Theta_{ij} \) is the directional cosine factor corresponding to \( \vec{v} \cdot \vec{l} / |\vec{v}| |\vec{l}| \) for the \( i \)-th measurement and \( j \)-th point.

To explicitly indicate that the data are observed and not random variables, superscripts with "obs" notation are introduced, as in the following equations:
\begin{eqnarray}
&&\bm{I}_0^\mathrm{obs} = \bm{g}_0(\hat{\bm{e}}) + \bm{\epsilon}_0,\label{eq: def_of_g0obs}\\ 
&&\bm{I}_\mathrm{Re}^\mathrm{obs} = \bm{g}_\mathrm{C}(\hat{\bm{a}},\hat{\bm{v}}) + \bm{\epsilon}_1,\quad 
\bm{I}_\mathrm{Im}^\mathrm{obs} =  \bm{g}_\mathrm{S}(\hat{\bm{a}},\hat{\bm{v}}) + \bm{\epsilon}_1, \label{eq: def_of_IRI_obs}
\end{eqnarray}
which are acquired from \( I_0 \), \( \mathrm{Re}(I_\mathrm{D}) \), and \( \mathrm{Im}(I_\mathrm{D}) \), respectively. 
Here, \( \bm{\epsilon}_0 \) and \( \bm{\epsilon}_1 \) represent measurement noise in the observations, assumed to be normally distributed with zero mean and covariances \( \Sigma_0 \) and \( \Sigma_1 \), respectively.

\section{Nonlinear Gaussian Process Tomography for Doppler Spectral Imaging}\label{sec:Nonlinear_GPT_CIS}

In this section, we solve the nonlinear CIS inverse problem: estimate the posterior of local ion temperature and velocity given the observed CIS images, while treating emissivity as a latent variable to be marginalized.

\subsection{Bayesian Framework for Tomography}

\begin{figure}[htbp]
    \centering
    \includegraphics{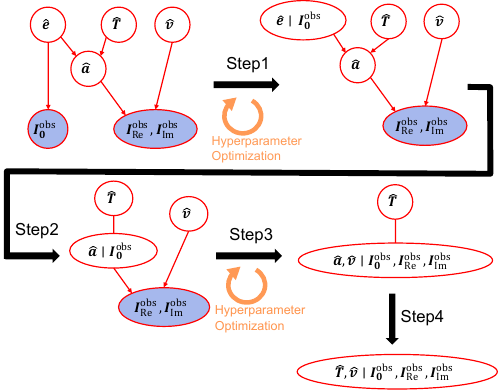}
    \caption{\label{fig: Bayesian_graphical_model2} Conceptual diagrams of the tomographic model for CIS. The first diagram is a  Bayesian graphical model of the CIS projection equations, consisting of Eqs.~\eqref{eq: def_of_a}, \eqref{eq: discretized_g0}, \eqref{eq: discretized_gc}, \eqref{eq: discretized_gs}, \eqref{eq: def_of_g0obs}, and \eqref{eq: def_of_IRI_obs}. Circles represent random variables, and blue shaded circles denote observed variables, following the graphical model notation used in the PRML~\cite{Bishop2006}. Step 1 computes the posterior probability of the log-emissivity \( \hat{\bm{e}} \) given the observed data \( \bm{I}_0^\mathrm{obs} \). Step 2 marginalizes the variable \( \hat{\bm{e}} \). Step 3 computes the posterior probabilities of \( \hat{\bm{a}} \) and \( \hat{\bm{v}} \). Step 4 marginalizes the variable \( \hat{\bm{a}} \).}
\end{figure}

In this framework, the variables \( \hat{e} \), \( \hat{T} \), and \( \hat{v} \) are modeled as Gaussian processes. Therefore, the discretized vectors of these variables follow multivariate Gaussian distributions, denoted as:
\begin{eqnarray*}
    p(\hat{\bm{e}}) &=& \mathcal{N}(\hat{\bm{e}} \mid \bm{\mu}^\mathrm{pri}_{e}, K_{e}),\\ 
    p(\hat{\bm{T}}) &=& \mathcal{N}(\hat{\bm{T}} \mid \bm{\mu}^\mathrm{pri}_T, K_T),\\
    p(\hat{\bm{v}}) &=& \mathcal{N}(\hat{\bm{v}} \mid \bm{\mu}^\mathrm{pri}_v, K_v),
\end{eqnarray*}
where \( \bm{\mu}^\mathrm{pri}_{e} \), \( \bm{\mu}^\mathrm{pri}_T \), and \( \bm{\mu}^\mathrm{pri}_v \) are the prior mean vectors of the log-emissivity, temperature, and velocity, respectively, and \( K_{e} \), \( K_T \), \( K_v \) are the prior covariance matrices.

The goal of Bayesian estimation in this section is to find the posterior probability of the local variables \( \hat{\bm{T}} \) and \( \hat{\bm{v}} \) given the observed data from the CIS signals, yielding \( p(\hat{\bm{T}}, \hat{\bm{v}} \mid \bm{I}_{0}^\mathrm{obs}, \bm{I}_\mathrm{Re}^\mathrm{obs}, \bm{I}_\mathrm{Im}^\mathrm{obs}) \). 
Although the variable \( \hat{\bm{e}} \) seems unrelated to the primary objective, it significantly influences the tomography for temperature and velocity by mediating through the local amplitude \( \hat{\bm{a}} \), as defined in Eq.~\eqref{eq: def_of_a}.
Therefore, all three equations (Eqs.~\eqref{eq: discretized_g0}, \eqref{eq: discretized_gc}, \eqref{eq: discretized_gs}) must be considered, and then the unnecessary variables \( \hat{\bm{e}} \) and \( \hat{\bm{a}} \) will be marginalized later, as described below:
\begin{eqnarray}
&& p(\hat{\bm{T}}, \hat{\bm{v}} \mid \bm{I}_0^\mathrm{obs}, \bm{I}_\mathrm{Re}^\mathrm{obs}, \bm{I}_\mathrm{Im}^\mathrm{obs})\nonumber\\ &=& \iint p(\hat{\bm{T}}, \hat{\bm{v}}, \hat{\bm{a}}, \hat{\bm{e}} \mid \bm{I}_0^\mathrm{obs}, \bm{I}_\mathrm{Re}^\mathrm{obs}, \bm{I}_\mathrm{Im}^\mathrm{obs}) \, \mathrm{d}\hat{\bm{a}} \, \mathrm{d}\hat{\bm{e}} \nonumber\\
&=& \iint p(\hat{\bm{a}}, \hat{\bm{v}} \mid \bm{I}_\mathrm{Re}^\mathrm{obs}, \bm{I}_\mathrm{Im}^\mathrm{obs}) \delta(\hat{\bm{T}} - \hat{\bm{e}} + \hat{\bm{a}}) p(\hat{\bm{e}} \mid \bm{I}_0^\mathrm{obs}) \, \mathrm{d}\hat{\bm{a}} \, \mathrm{d}\hat{\bm{e}},\nonumber\\
\label{eq: marginalization}
\end{eqnarray}
where \( \delta(\cdot) \) is the Dirac delta function.

Since the straightforward calculation of Eq.~\eqref{eq: marginalization} is still too complicated to solve, the following step-by-step procedure is presented instead.
 A graphical representation of this procedure is shown in Fig.~\ref{fig: Bayesian_graphical_model2}.

\begin{itemize}
    \item \textbf{Step 1.} Estimate the posterior of log-emissivity from \( \bm{I}_0^\mathrm{obs} \) using the Laplace approximation:
    \begin{eqnarray}
        p(\hat{\bm{e}} \mid \bm{I}_0^\mathrm{obs}) 
        \overset{\mathrm{LA}}{\simeq} \mathcal{N}(\hat{\bm{e}} \mid \tilde{\bm{\mu}}_{e}^\mathrm{LA}, \tilde{\Sigma}_{e}^\mathrm{LA}).\label{eq: approximate_posterior_of_emissivity}
    \end{eqnarray}
    
    \item \textbf{Step 2.} Introduce the mediating variable \( \hat{\bm{a}} \) with \( p(\hat{\bm{a}} \mid \hat{\bm{e}}, \hat{\bm{T}}) = \delta(\hat{\bm{a}} - \hat{\bm{e}} + \hat{\bm{T}}) \), and marginalize \( \hat{\bm{e}} \) to obtain the prior:
    \begin{eqnarray}
    p(\hat{\bm{a}} \mid \bm{I}_0^\mathrm{obs}) = \mathcal{N}(\hat{\bm{a}} \mid \tilde{\bm{\mu}}_{e}^\mathrm{LA} - \bm{\mu}^\mathrm{pri}_T, \tilde{\Sigma}^\mathrm{LA}_{e} + K_T). \label{eq: posterior_a}
    \end{eqnarray}
    We denote \( \bm{\mu}_a^\mathrm{pri} := \tilde{\bm{\mu}}_{e}^\mathrm{LA} - \bm{\mu}^\mathrm{pri}_T \) and \( \Sigma_{a}^\mathrm{pri} := \tilde{\Sigma}^\mathrm{LA}_{e} + K_T \).
    
    \item \textbf{Step 3.} Using the priors \( \hat{\bm{a}} \sim \mathcal{N}(\bm{\mu}_a^\mathrm{pri}, \Sigma_a^\mathrm{pri}) \) and \( \hat{\bm{v}} \sim \mathcal{N}(\bm{\mu}_v^\mathrm{pri}, K_v) \), compute the joint posterior via Laplace approximation:
    \begin{eqnarray*}
    p(\hat{\bm{a}}, \hat{\bm{v}} \mid \bm{I}_0^\mathrm{obs}, \bm{I}_\mathrm{Re}^\mathrm{obs}, \bm{I}_\mathrm{Im}^\mathrm{obs}) \overset{\mathrm{LA}}{\simeq} 
    \mathcal{N} \left(
    \begin{pmatrix}
        \hat{\bm{a}} \\
        \hat{\bm{v}}
    \end{pmatrix}
    \Bigg| 
    \begin{pmatrix}
        \tilde{\bm{\mu}}^\mathrm{LA}_{a} \\
        \tilde{\bm{\mu}}^\mathrm{LA}_{v}
    \end{pmatrix}
    ,
    \begin{bmatrix}
        \tilde{\Sigma}^\mathrm{LA}_{aa} & \tilde{\Sigma}^\mathrm{LA}_{av}\\
        \tilde{\Sigma}^\mathrm{LA}_{va} & \tilde{\Sigma}^\mathrm{LA}_{vv}
    \end{bmatrix}
    \right).
    \end{eqnarray*} 
    
    \item \textbf{Step 4.} Marginalize \( \hat{\bm{a}} \) to obtain the posterior of \( \hat{\bm{T}} \) and \( \hat{\bm{v}} \):
    \begin{eqnarray*}
        p(\hat{\bm{T}}, \hat{\bm{v}} \mid \bm{I}_0^\mathrm{obs}, \bm{I}_\mathrm{Re}^\mathrm{obs}, \bm{I}_\mathrm{Im}^\mathrm{obs}) =  
        \mathcal{N} \left(
            \begin{pmatrix}
                \hat{\bm{T}} \\
                \hat{\bm{v}}
            \end{pmatrix}
            \Bigg| 
        \begin{pmatrix}
            \tilde{\bm{\mu}}_{T} \\
            \tilde{\bm{\mu}}_{v}
        \end{pmatrix}
        ,
        \begin{bmatrix}
            \tilde{\Sigma}_{TT} & \tilde{\Sigma}_{Tv}\\
            \tilde{\Sigma}_{vT} & \tilde{\Sigma}_{vv}
        \end{bmatrix}
        \right),
    \end{eqnarray*}
    with mean vectors and covariance matrices:
    \begin{eqnarray}
        \tilde{\bm{\mu}}_{T} &=& \bm{\mu}^\mathrm{pri}_{T} + K_T [\Sigma^\mathrm{pri}_a]^{-1} (\tilde{\bm{\mu}}^\mathrm{LA}_{e} - \bm{\mu}^\mathrm{pri}_{T} - \tilde{\bm{\mu}}^\mathrm{LA}_{a}),\nonumber\\
        \tilde{\bm{\mu}}_{v} &=& \tilde{\bm{\mu}}^\mathrm{LA}_{v}, \nonumber\\
        \tilde{\Sigma}_{TT} &=& K_T [\Sigma^\mathrm{pri}_a]^{-1} \tilde{\Sigma}^\mathrm{LA}_{e}
         + K_T [\Sigma^\mathrm{pri}_a]^{-1} \tilde{\Sigma}_{aa}^\mathrm{LA} [\Sigma^\mathrm{pri}_a]^{-1} K_T,\nonumber\\
        \tilde{\Sigma}_{Tv} &=& -K_T [\Sigma^\mathrm{pri}_a]^{-1} \tilde{\Sigma}_{av}^\mathrm{LA} = \tilde{\Sigma}_{vT}^\mathrm{T},\nonumber\\
        \tilde{\Sigma}_{vv} &=& \tilde{\Sigma}^\mathrm{LA}_{vv}.\label{eq: transfer_T_from_a}
    \end{eqnarray}
\end{itemize}

\subsection{Laplace Approximation}\label{subsec: Laplace_approximation}

As shown in Step 1 and Step 3 of the previous subsection, the Laplace approximation is used to compute the posterior probability of the local variables \( \hat{\bm{e}} \), \( \hat{\bm{a}} \), and \( \hat{\bm{v}} \) in the CIS model.

In the case of the log-emissivity \( \hat{\bm{e}} \), the log-posterior function \( \Psi^\mathrm{emit}(\hat{\bm{e}}) \) is defined using Eqs.~\eqref{eq: discretized_g0} and \eqref{eq: def_of_g0obs} as follows:
\begin{eqnarray*}
    \Psi^\mathrm{emit}(\hat{\bm{e}}) &=& \log{p(\hat{\bm{e}} \mid \bm{I}_0^\mathrm{obs}) } + C\\
    &=& -\frac{1}{2}(\bm{g}_0(\hat{\bm{e}}) - \bm{I}_0^\mathrm{obs})^\mathrm{T} \Sigma_{g_0}^{-1} (\bm{g}_0(\hat{\bm{e}}) - \bm{I}_0^\mathrm{obs}) \\
    && -\frac{1}{2}(\hat{\bm{e}} - \bm{\mu}^\mathrm{pri}_e)^\mathrm{T} K_{e}^{-1} (\hat{\bm{e}} - \bm{\mu}^\mathrm{pri}_e),
\end{eqnarray*}
where \( \Sigma_{g_0} \) is the covariance matrix of the measurement error \( \bm{\epsilon}_{0} \) described in Eq.~\eqref{eq: def_of_g0obs}, and \( C \) is a constant term that does not depend on \( \hat{\bm{e}} \).

Using matrix calculus, the analytical expressions of the gradient \( \bm{\nabla} \Psi^\mathrm{emit}(\hat{\bm{e}}) \) and the Hessian \( \nabla^2 \Psi^\mathrm{emit}(\hat{\bm{e}}) \) are derived as follows:
\begin{eqnarray}
  \{\bm{\nabla} \Psi^\mathrm{emit}(\hat{\bm{e}})\}_i
  &=& \left\{ H^\mathrm{T} \Sigma_{g_0}^{-1} (\bm{g}_0(\hat{\bm{e}}) - \bm{I}_{0}^\mathrm{obs}) \right\}_i
   \exp{\hat{e}_{i}} \nonumber\\
  &-& \left\{ K_{e}^{-1} (\hat{\bm{e}} - \bm{\mu}_{e}^\mathrm{pri}) \right\}_i,\label{eq: nabla_psi_e} \\
  \{\nabla^2 \Psi^\mathrm{emit}(\hat{\bm{e}})\}_{ij}
  &=& \left[ H^\mathrm{T}  \Sigma_{g_0}^{-1} H \right]_{ij}
  \exp{\hat{e}_{i}} \exp{\hat{e}_{j}} \nonumber\\
   &-& \delta_{ij} \left\{ H^\mathrm{T}  \Sigma_{g_0}^{-1} (\bm{g}_0(\hat{\bm{e}}) - \bm{I}_{0}^\mathrm{obs}) \right\}_i
   \exp{\hat{e}_{i}} \nonumber\\
  &-& \left\{ K_{e}^{-1} \right\}_{ij}, \label{eq: laplacian_psi_e}  
\end{eqnarray}
where \( i, j \) are indices of the vectors or matrices, and \( \delta_{ij} \) is the Kronecker delta.
By substituting Eqs.~\eqref{eq: nabla_psi_e} and \eqref{eq: laplacian_psi_e} into Eq.~\eqref{eq: Newtonmethod} and iterating until convergence is achieved, we use Eqs.~\eqref{eq: def_of_LA_mu} and \eqref{eq: def_of_LA_Sigma} to obtain the approximate mean vector \( \tilde{\bm{\mu}}^\mathrm{LA}_e \) and the approximate covariance matrix \( \tilde{\Sigma}^\mathrm{LA}_e \) of the posterior probability.

For the local amplitude \( \hat{\bm{a}} \) and local velocity \( \hat{\bm{v}} \), the log-posterior function \( \Psi^\mathrm{CIS}(\hat{\bm{a}}, \hat{\bm{v}}) \) is described using Eqs.~\eqref{eq: discretized_gc}, \eqref{eq: discretized_gs}, and \eqref{eq: def_of_IRI_obs} as follows:
\begin{eqnarray*}
    \Psi^\mathrm{CIS}(\hat{\bm{a}}, \hat{\bm{v}}) &:=& \log{p(\hat{\bm{a}}, \hat{\bm{v}} \mid \bm{I}_\mathrm{Re}^\mathrm{obs}, \bm{I}_\mathrm{Im}^\mathrm{obs})} - C \\
    &=& -\frac{1}{2}(\bm{g}_\mathrm{C}(\hat{\bm{a}}, \hat{\bm{v}}) - \bm{I}_\mathrm{Re}^\mathrm{obs})^\mathrm{T} \Sigma_{g_1}^{-1} (\bm{g}_\mathrm{C}(\hat{\bm{a}}, \hat{\bm{v}}) - \bm{I}_\mathrm{Re}^\mathrm{obs}) \\
    && -\frac{1}{2} (\bm{g}_\mathrm{S}(\hat{\bm{a}}, \hat{\bm{v}}) - \bm{I}_\mathrm{Im}^\mathrm{obs})^\mathrm{T} \Sigma_{g_1}^{-1} (\bm{g}_\mathrm{S}(\hat{\bm{a}}, \hat{\bm{v}}) - \bm{I}_\mathrm{Im}^\mathrm{obs}) \\
    && -\frac{1}{2}(\hat{\bm{a}} - \bm{\mu}^\mathrm{pri}_a)^\mathrm{T} (\Sigma_{a}^\mathrm{pri})^{-1} (\hat{\bm{a}} - \bm{\mu}^\mathrm{pri}_a) \\
    && -\frac{1}{2}(\hat{\bm{v}} - \bm{\mu}^\mathrm{pri}_v)^\mathrm{T} K_{v}^{-1} (\hat{\bm{v}} - \bm{\mu}^\mathrm{pri}_v),
\end{eqnarray*}
where \( \Sigma_{g_1} \) is the covariance matrix of the observation error \( \bm{\epsilon}_{1} \) described in Eq.~\eqref{eq: def_of_IRI_obs}, and \( C \) is a constant term that does not depend on \( \hat{\bm{a}} \) and \( \hat{\bm{v}} \).

The calculation procedure is analogous to that for the log-emissivity, but it should be noted that the total size of variables is doubled because joint probabilities of \( \hat{\bm{a}} \) and \( \hat{\bm{v}} \) are considered. In this case, we define the gradient and Hessian of the log-posterior function \( \Psi^\mathrm{CIS}(\hat{\bm{a}}, \hat{\bm{v}}) \) as block matrices:
\begin{eqnarray}
  \bm{\nabla} \Psi^{\mathrm{CIS}}(\hat{\bm{a}}, \hat{\bm{v}}) &=&
      \begin{bmatrix}
        \frac{\partial}{\partial \hat{\bm{a}}}  \\
        \frac{\partial}{\partial \hat{\bm{v}}}\\
      \end{bmatrix}
      \Psi^{\mathrm{CIS}}(\hat{\bm{a}}, \hat{\bm{v}}),\label{eq: def_of_nabla_Psi^CIS}\\
  \nabla^2\Psi^{\mathrm{CIS}}(\hat{\bm{a}}, \hat{\bm{v}}) &=&
      \begin{bmatrix}
        \frac{\partial^2}{\partial \hat{\bm{a}}\partial \hat{\bm{a}}} & \frac{\partial^2}{\partial \hat{\bm{a}}\partial \hat{\bm{v}}} \\
        \frac{\partial^2}{\partial \hat{\bm{v}}\partial \hat{\bm{a}}} & \frac{\partial^2}{\partial \hat{\bm{v}}\partial \hat{\bm{v}}} \\
      \end{bmatrix}
      \Psi^{\mathrm{CIS}}(\hat{\bm{a}}, \hat{\bm{v}}).\label{eq: def_of_Lap_Psi^CIS}
\end{eqnarray}
Due to the presence of the directional cosine matrix \( \Theta \), the specifics of Eqs.~\eqref{eq: def_of_nabla_Psi^CIS} and \eqref{eq: def_of_Lap_Psi^CIS} are complex and are detailed in Appendix~\ref{app:derive_Phi_cis}. The components of the gradient are given in Eqs.~\eqref{eq: dPsida_appendix} and \eqref{eq: dPsida_appendix}, and the components of the Hessian are given in Eqs.~\eqref{eq: ddPsidada_appendix}, \eqref{eq: ddPsidvdv_appendix}, and \eqref{eq: ddPsidadv_appendix}.

\subsection{Hyperparameter Optimization}

In the Bayesian framework, the hyperparameters of the Gaussian process priors, such as the prior mean vectors and covariance matrices (\( \bm{\mu}^\mathrm{pri}_{e} \), \( K_{e} \), \( \bm{\mu}^\mathrm{pri}_{T} \), \( K_{T} \), \( \bm{\mu}^\mathrm{pri}_{v} \), \( K_{v} \)), as well as the observation noise covariance matrices (\( \Sigma_{g_0} \), \( \Sigma_{g_1} \)), play a crucial role in the performance of the tomography.
To determine appropriate values for these hyperparameters, we use the evidence approximation obtained via the Laplace approximation. By maximizing the approximate log-marginal likelihood with respect to the hyperparameters, we find their optimal values that best explain the observed data.

For the emission model, the approximate log-marginal likelihood \( \mathcal{L}^\mathrm{emit} \) is given by:
\begin{eqnarray}
\mathcal{L}^\mathrm{emit}(\bm{\theta}) &=&
\Psi^\mathrm{emit}(\tilde{\bm{\mu}}^\mathrm{LA}_{e}) - \frac{1}{2}\log \lvert K_{e} \rvert - \frac{1}{2}\log \lvert \Sigma_{g_0} \rvert \nonumber\\
&& + \frac{1}{2}\log \lvert \tilde{\Sigma}_{e}^\mathrm{LA}\rvert, \label{eq: evidence_for_emission}
\end{eqnarray}
where $\bm{\theta}$ collectively denotes the hyperparameters of the emissivity model, including $\bm{\mu}_{e}^{\mathrm{pri}}$, $K_{e}$, and $\Sigma_{g_0}$.
In this equation, \( \Psi^\mathrm{emit}(\tilde{\bm{\mu}}^\mathrm{LA}_{e}) \) is the log-posterior function evaluated at the Laplace approximation mean \( \tilde{\bm{\mu}}^\mathrm{LA}_{e} \).

Similarly, for the CIS model, the approximate log-marginal likelihood \( \mathcal{L}^\mathrm{CIS} \) is given by:
\begin{eqnarray}
\mathcal{L}^{\mathrm{CIS}}(\bm{\theta}) &=& \Psi^\mathrm{CIS}(\tilde{\bm{\mu}}^\mathrm{LA}_{a}, \tilde{\bm{\mu}}^\mathrm{LA}_{v}) 
- \frac{1}{2}\log \lvert \tilde{\Sigma}^\mathrm{LA}_{e} + K_T  \rvert - \frac{1}{2}\log \lvert K_{v} \rvert \nonumber \\
&& - \log \lvert \Sigma_{g_1} \rvert + \frac{1}{2}\log
\left|
\begin{array}{cc}
    \tilde{\Sigma}^\mathrm{LA}_{aa} & \tilde{\Sigma}^\mathrm{LA}_{av}\\
    \tilde{\Sigma}^\mathrm{LA}_{va} & \tilde{\Sigma}^\mathrm{LA}_{vv}
\end{array}
\right|, \label{eq: evidence_for_CIS}
\end{eqnarray}
where $\bm{\theta}$ collectively denotes the hyperparameters of the CIS model, including $\bm{\mu}_{a}^{\mathrm{pri}}$, $\bm{\mu}_{v}^{\mathrm{pri}}$, $\Sigma_{a}^{\mathrm{pri}}$, $K_{v}$, and $\Sigma_{g_1}$. 
Here, \( \Psi^\mathrm{CIS}(\tilde{\bm{\mu}}^\mathrm{LA}_{a}, \tilde{\bm{\mu}}^\mathrm{LA}_{v}) \) is the log-posterior function evaluated at the Laplace approximation means \( \tilde{\bm{\mu}}^\mathrm{LA}_{a} \) and \( \tilde{\bm{\mu}}^\mathrm{LA}_{v} \).

By maximizing \( \mathcal{L}^\mathrm{emit}(\bm{\theta}) \) and \( \mathcal{L}^\mathrm{CIS}(\bm{\theta}) \) with respect to the hyperparameters \( \bm{\theta} \), we can obtain their optimal values.
This procedure ensures that the Gaussian process models for the log-emissivity, temperature, and velocity are appropriately tuned to the observed data, improving the accuracy and reliability of the tomographic reconstructions.
\section{Verification with synthetic Phantom data}\label{sec:Test_Phantom_Data}

\begin{figure*}[htbp]
    \centering
    \includegraphics{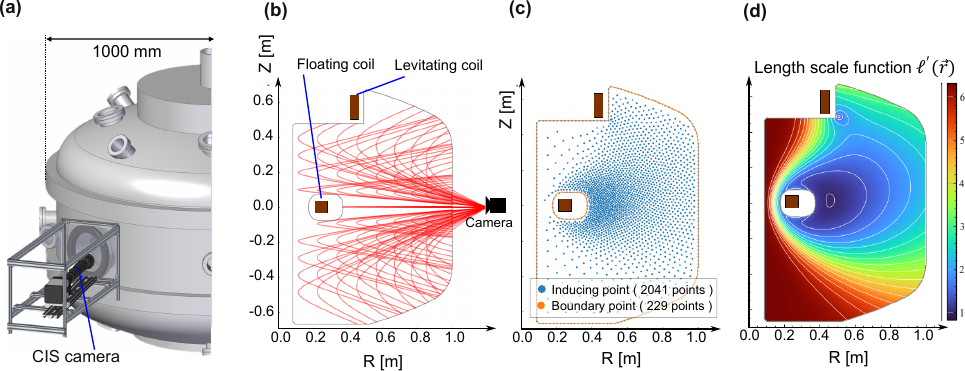}
    \caption{\label{fig: RT-1_configuration} (a) A conceptual diagram of the camera system in RT-1, where the CIS camera is installed tangentially to the plasma cross-section to detect toroidal flow. (b) Examples of projected rays on the poloidal cross-section of RT-1. The number of rays is reduced for simplicity. (c) The distribution of scattered inducing points \( \vec{r}^\mathrm{idc} \) (the blue points) and boundary points \( \vec{r}^\mathrm{bd} \) (the orange points). The number of inducing points is 2041, and the number of boundary points is 229 in the setup. (d) The length scale function \( \ell'(\vec{r}) \) in RT-1. These conditions are consistent with previous work~\cite{Ueda2025-pc}.}  
\end{figure*}

\begin{figure*}[htb]
    \centering
      \includegraphics{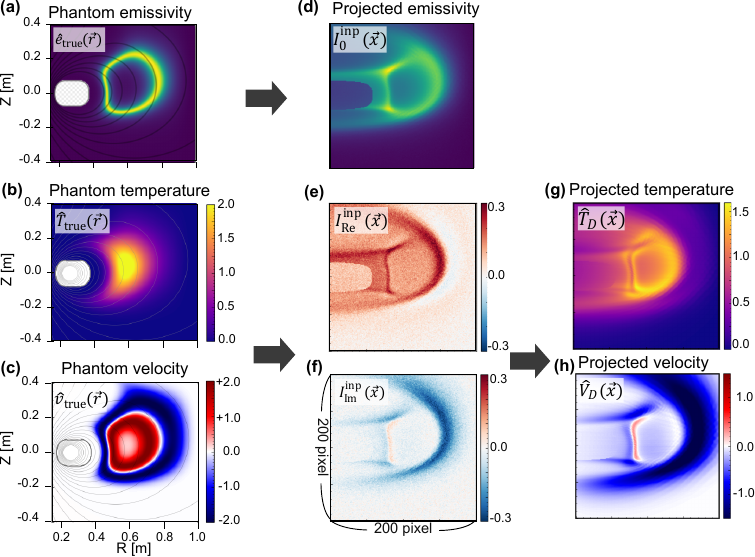}
      \caption{\label{fig: phantom_cis_GPT} Phantom data for the test tomography and the corresponding projected images. The left column, (a), (b), and (c), are phantom distributions of local emissivity (\(e_\mathrm{true}\)), local temperature (\(\hat{T}_\mathrm{true}\)), and local velocity (\(\hat{v}_\mathrm{true}\)), respectively. Panel (d) is the projected emissivity, \(\bm{I}_0^\mathrm{inp}\), using Eqs.~\eqref{eq: discretized_g0} and~\eqref{eq: def_of_g0obs}. Panels (e) and (f) are input images for CIS tomography generated by Eqs.~\eqref{eq: discretized_gc},~\eqref{eq: discretized_gs}, and~\eqref{eq: def_of_IRI_obs}, corresponding to \(\bm{I}_\mathrm{Re}^\mathrm{inp}\) and \(\bm{I}_\mathrm{Im}^\mathrm{inp}\), respectively. Panels (g) and (h) are the projected temperature and projected velocity, respectively.}
\end{figure*}

\begin{figure}[htb]
    \centering
      \includegraphics{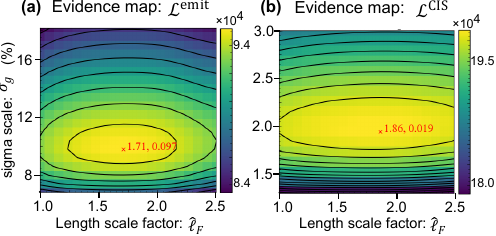}
        \caption{\label{fig: evidence_map} Evidence maps for the CIS tomography model when the input images are shown in Fig.~\ref{fig: phantom_cis_GPT}. 
         (a) \(\mathcal{L}^\mathrm{emit}\) and (b) \(\mathcal{L}^\mathrm{CIS}\) are defined in Eqs.~\eqref{eq: evidence_for_emission} and~\eqref{eq: evidence_for_CIS}, respectively. The horizontal and vertical axes represent the length scale factor \(\hat{\ell}_F\) and the sigma scale \(\sigma_{g}\), which are defined in Eqs.~\eqref{eq: length_scale} and~\eqref{eq: sigma_g}, respectively. The red crosses indicate the maximum points.}
\end{figure}

\begin{figure*}[htbp]
    \centering
    \includegraphics{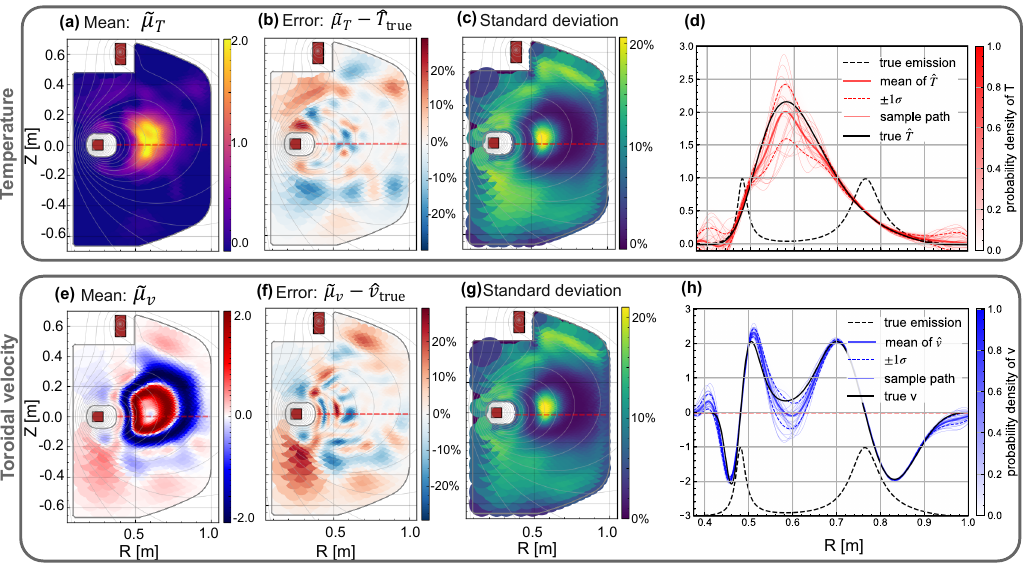} 
    \caption{\label{fig: phantom_result} The tomographic results of temperature and velocity with \(\sigma_{g} = 0.019\) and \(\hat{\ell}_{F} = 1.86\) when the phantom distributions and input images are shown in Fig.~\ref{fig: phantom_cis_GPT}. The top row represents the temperature, and the bottom row represents the velocity. The first column, (a) and (e), shows the posterior means of each variable. The second column, (b) and (f), displays the errors given by the mean minus the true value. The third column, (c) and (g), presents the standard deviations of the posteriors. The rightmost column, (d) and (h), shows the radial profiles at \( z = 0 \) m, where the black solid lines are the true values of each variable, black dashed lines are the true values of the emissivity, red and blue solid lines are the means of temperature and velocity, respectively, red and blue dashed lines are the \(-\sigma\) and \(+\sigma\) regions of the posterior probabilities of each variable, and thin colored lines are the sample paths of each posterior probability.}
\end{figure*}

\subsection{Configuration of the Tomographic Model for RT-1}\label{subsec: RT-1_configuration}

In this study, we simulate observations of the plasma in the RT-1 experimental device~\cite{Z_Yoshida_2010} using CIS~\cite{Nakamura_2018}.  Unlike confinement devices such as tokamaks and stellarators, RT-1 has only a poloidal magnetic field and achieves pure axisymmetry. In this sense, RT-1 is more suitable for reconstruction from a single image. The CIS installed in RT-1 is shown in Fig.~\ref{fig: RT-1_configuration}(a), with a tangential field of view to observe the toroidal flow. 
In this configuration, only the toroidal component of the flow is assumed.

The trajectories of the rays emanating from the camera are shown in Fig.~\ref{fig: RT-1_configuration}(b). The positions of the local variables \( f \) are defined as \( \bm{f} = f(\vec{\bm{r}}) \). In both conventional tomography methods and traditional GPT, it is common to arrange each of the \( \{\vec{r}_i\}_{i=1}^N \) on a grid. However, in our method, as shown in Fig.~\ref{fig: RT-1_configuration}(c), we arrange the points in a scatter plot that is not restricted to a grid shape. In Gaussian processes naturally defined in function space, it is not necessary to arrange the points on a grid. Moreover, by varying the density of the point cloud, we can reduce the dimensionality of \( \bm{f} \) and decrease the computational cost. The reason for adopting such an arrangement is that, in the RT-1 plasma, the local structure becomes more complex closer to the levitated coil within the magnetic surfaces, requiring higher resolution in those regions.

Regarding the length scale for the kernel function, we adopt an isotropic but non-uniform one. To adjust the parameters, we express \( \ell(\vec{r}) \) to substitute into the Gibbs kernel [Eq.~\eqref{eq: Gibbs_kernel}] as follows:
\begin{eqnarray}\label{eq: length_scale}
    \ell(\vec{r}) = \hat{\ell}_F \ell'(\vec{r}),
\end{eqnarray}
where \( \ell'(\vec{r}) \) is a non-uniform function given in Fig.~\ref{fig: RT-1_configuration}(d), and \( \hat{\ell}_F \) is the length scale factor chosen to maximize the evidence. However, since the distance intervals of the point cloud in Fig.~\ref{fig: RT-1_configuration}(c) correlate with \( \ell(\vec{r}) \), too small a length scale would result in insufficient degrees of freedom. Therefore, we impose the constraint \( \hat{\ell}_F > 1 \) during the optimization.

As boundary conditions, we define the boundary local variable vector \( \bm{f}^\mathrm{bd} = f(\vec{\bm{r}}^\mathrm{bd}) \) from the orange point set in Fig.~\ref{fig: RT-1_configuration}(c), and update the prior distribution \( \bm{\mu}_{f}^\mathrm{pri} \) and \( K_{f}^\mathrm{pri} \) as the conditional probability given the values at \( \bm{f}^\mathrm{bd} \). Here, we set the values of temperature \( \hat{T} \) and velocity \( \hat{v} \) at the boundary to zero, and for the log-emissivity \( \hat{e} \), we assign values in the range of \(-5\) to \(-3\).

Regarding \( \Sigma_{g_0} \) and \( \Sigma_{g_1} \) necessary to define the likelihood function, in this test, we use artificial noise assumed to be uniform white Gaussian noise. Therefore, we assume an identity matrix scaled by the noise variance:
\begin{eqnarray}\label{eq: sigma_g}
    \Sigma_{g} = \sigma_{g}^2  I,
\end{eqnarray}
where \( \sigma_{g} \) is the sigma scale chosen to maximize the evidence, similar to \( \hat{\ell}_F \).

\subsection{Synthetic Phantom Data}\label{subsec: phantom_data}

The phantom distributions of emissivity, temperature, and velocity are shown in panels (a), (b), and (c) of Fig.~\ref{fig: phantom_cis_GPT}, denoted as \( e_\mathrm{true} \), \( \hat{T}_\mathrm{true} \), and \( \hat{v}_\mathrm{true} \), respectively. These distributions are based on the assumption that the plasma emits in a ring shape, with high temperature inside the ring, and the velocity switches between positive and negative at the boundary of the ring. Note that these patterns are not directly related to the actual observations in RT-1.

The input data for the test are generated using Eqs.~\eqref{eq: discretized_g0} and~\eqref{eq: def_of_g0obs} for the emissivity (panel d), and Eqs.~\eqref{eq: discretized_gc}, \eqref{eq: discretized_gs}, and~\eqref{eq: def_of_IRI_obs} for the real and imaginary components (panels e and f), with the phantom distributions as inputs. The results are shown in panels (d), (e), and (f) of Fig.~\ref{fig: phantom_cis_GPT}, denoted as \( \bm{I}_0^\mathrm{inp} \), \( \bm{I}_\mathrm{Re}^\mathrm{inp} \), and \( \bm{I}_\mathrm{Im}^\mathrm{inp} \), respectively.

Panels (g) and (h) in Fig.~\ref{fig: phantom_cis_GPT} represent the projected temperature and velocity, which are calculated by~\cite{Howard_2003}:
\begin{eqnarray}\label{eq: T_D_and_v_D}
    \hat{T}_D = \log \sqrt{\frac{I_0^2}{I_\mathrm{Re}^2 + I_\mathrm{Im}^2}},\quad
    \hat{v}_D = \arctan \frac{I_\mathrm{Im}}{I_\mathrm{Re}},
\end{eqnarray}
and correspond to \( \log \zeta_{D} \) and \( \phi_D \) in Eq.~\eqref{eq: signals of CIS}, respectively.

In contrast to the relationship between the emissivity of the projected image Fig.~\ref{fig: phantom_cis_GPT}(d) and the phantom distribution Fig.~\ref{fig: phantom_cis_GPT}(a), Figs.~\ref{fig: phantom_cis_GPT}(g) and~\ref{fig: phantom_cis_GPT}(h) indicate that for temperature and velocity, it is difficult for humans to infer the original distributions from the projected images. This is because the three physical variables influence each other through the integration process, resulting in more complex outputs. Additionally, we emphasize that the peak values of both the normalized temperature \( \hat{T} \) and normalized velocity \( \hat{v} \) are set to be around 2. These values are sufficiently high that linear approximations are invalid.

To validate the tomography, artificial noise is added to the generated projection images as input data. The noise is uniform white Gaussian noise, and the noise level is set to 10\% for \( \bm{I}_0 \) and 2\% for \( \bm{I}_\mathrm{Re} \) and \( \bm{I}_\mathrm{Im} \). The noise level is defined as the ratio of the standard deviation of the white noise to the mean of \( \bm{I}_0 \).

\subsection{Results of the Test}

Based on the hyperparameters \( \bm{\mu}_{e}^\mathrm{pri}, K_{e} \), \( \Sigma_{g_0} \), \( \bm{\mu}_{T}^\mathrm{pri}\), \(K_{T} \), \( \bm{\mu}_{v}^\mathrm{pri}\), \(K_{v} \), and \( \Sigma_{g_1} \), and input images \( \bm{I}_0^\mathrm{inp}, \bm{I}_\mathrm{Re}^\mathrm{inp}, \bm{I}_\mathrm{Im}^\mathrm{inp} \), we perform the tomography using the nonlinear GPT for CIS described in Sec.~\ref{sec:Nonlinear_GPT_CIS}. For hyperparameter optimization, there are two objective functions, \( \mathcal{L}^\mathrm{emit} \) and \( \mathcal{L}^\mathrm{CIS} \), which are given in Eqs.~\eqref{eq: evidence_for_emission} and~\eqref{eq: evidence_for_CIS}, respectively. Before proceeding to Step 2, \( \mathcal{L}^\mathrm{emit} \) is optimized, and the optimal \( \tilde{\bm{\mu}}_{e}^\mathrm{LA} \) and \( \tilde{\Sigma}_{e}^\mathrm{LA} \) are used in Step 2. Then, the optimization of \( \mathcal{L}^\mathrm{CIS} \) is performed in Step 3. Finally, we obtain the mean vectors \( \tilde{\bm{\mu}}_T \) and \( \tilde{\bm{\mu}}_v \), and covariance matrices \( \tilde{\Sigma}_{TT} \), \( \tilde{\Sigma}_{Tv} \), \( \tilde{\Sigma}_{vT} \), and \( \tilde{\Sigma}_{vv} \).

Figure~\ref{fig: evidence_map} shows the evidence maps of \( \mathcal{L}^\mathrm{emit} \) and \( \mathcal{L}^\mathrm{CIS} \) with respect to the length scale factor \( \hat{\ell}_F \) and the sigma scale \( \sigma_{g} \). The optimal values are obtained at \( (\hat{\ell}_F, \sigma_{g_0}) = (1.71, 9.7\%) \) for the emissivity model, and at \( (\hat{\ell}_F, \sigma_{g_1}) = (1.86, 1.9\%) \) for the CIS model. Regarding the sigma scale, the optimal values are almost the same as the noise levels, which are 10\% for \( \bm{I}_0 \) and 2\% for \( \bm{I}_\mathrm{Re} \) and \( \bm{I}_\mathrm{Im} \), indicating that the noise levels can be predicted by maximizing the evidence when the error follows white Gaussian noise.

Tomographic results with optimal hyperparameters are shown in Fig.~\ref{fig: phantom_result}, where the distributions of temperature and velocity on the poloidal cross-section of RT-1 are displayed. The mean values of the posterior temperature and velocity, \( \tilde{\bm{\mu}}_T \) and \( \tilde{\bm{\mu}}_v \), are shown in panels (a) and (e) of Fig.~\ref{fig: phantom_result}, respectively. Error values, calculated as \( \tilde{\bm{\mu}}_T - \hat{T}_\mathrm{true} \) and \( \tilde{\bm{\mu}}_v - \hat{v}_\mathrm{true} \), are shown in panels (b) and (f), respectively. Standard deviations of the posterior temperature and velocity, given by
\begin{eqnarray*}
    \{\tilde{\sigma}_{T}\}_i = [\tilde{\Sigma}_{TT}]_{ii}^{1/2}, \quad
    \{\tilde{\sigma}_{v}\}_i = [\tilde{\Sigma}_{vv}]_{ii}^{1/2},
\end{eqnarray*}
are shown in panels (c) and (g), respectively.

Regions with large standard deviations of temperature and velocity correspond to regions of low emissivity, which is consistent with the property in CIS that spectral information is not propagated to the measurement in the absence of emissivity. Comparing the distributions of errors [(b), (f)] with the distributions of standard deviations [(c), (g)], the standard deviations roughly envelop the errors, which implies that the errors are predictable from the variance of the posterior probabilities.

These trends are also confirmed by the radial profiles at \( z = 0 \) m, as shown in panels (d) and (h) of Fig.~\ref{fig: phantom_result}, which show that the temperature and velocity deviate from the true values due to low emission in the range of \( 0.5\,\mathrm{m} < R < 0.7\,\mathrm{m} \), but this is compensated for by the increase in the confidence interval.
\section{Application to CIS measurements in RT-1}\label{sec:Tomography_Experimental_Data}

\begin{figure}[htbp]
    \centering
    \includegraphics{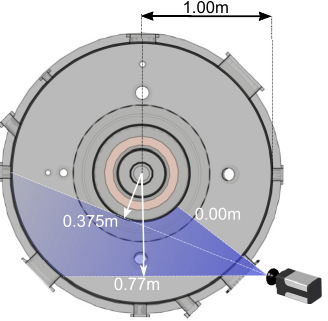}
    \caption{\label{fig: top_view_CIS} Top view of the CIS field of view on the equatorial plane in RT-1.}
\end{figure}

In this section, we present the tomographic reconstruction of ion temperature and velocity in the RT-1 device using data from the CIS diagnostic system. Details of the CIS system in RT-1 are described in Ref.~\cite{K_Ueda_2021}. The CIS cell used in this experiment has a \( \check{\phi}_0 \) of \( 124.1\,\mathrm{rad/nm} \), corresponding to a characteristic velocity \( v_\mathrm{c} \) of \( 5.17\,\mathrm{km/s} \) and a characteristic temperature \( T_\mathrm{c} \) of \( 2.23\,\mathrm{eV} \) for the He~II line (468.58~nm). Figure~\ref{fig: top_view_CIS} shows the field of view and installation of the CIS system.

\begin{figure}[htbp]
    \centering
    \includegraphics{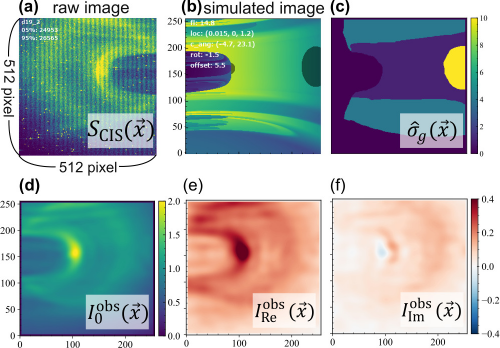}
    \caption{\label{fig: CIS_exp_images} Input images for CIS tomography. (a) Raw images measured by the CIS system. (b) Simulated image. (c) Relative sigma image \(\hat{\sigma}_g(\vec{x})\). (d), (e), and (f) are the input images derived by Fourier analysis: \( I_{0}^\mathrm{obs}(\vec{x}) \), \( I_\mathrm{Re}^\mathrm{obs}(\vec{x}) \), and \( I_\mathrm{Im}^\mathrm{obs}(\vec{x}) \), respectively.}
\end{figure}

The input images for the tomography model are shown in Fig.~\ref{fig: CIS_exp_images}. Panel~(a) shows the raw images, which were acquired three times under the same plasma discharge conditions with an exposure time of 0.5~s. Panel~(b) shows a simulated image used for adjusting the location, focal length, and angle of view of the camera.  Panels~(d), (e), and (f) are the input images derived by Fourier analysis and calibration techniques~\cite{K_Ueda_2021}, corresponding to \( I_{0}^\mathrm{obs}(\vec{x}) \), \( I_\mathrm{Re}^\mathrm{obs}(\vec{x}) \), and \( I_\mathrm{Im}^\mathrm{obs}(\vec{x}) \), respectively. These input images are reduced in size from \( 512 \times 512 \) pixels to \( 256 \times 256 \) pixels for computational reasons.

\begin{figure*}[htbp]
    \centering
    \includegraphics{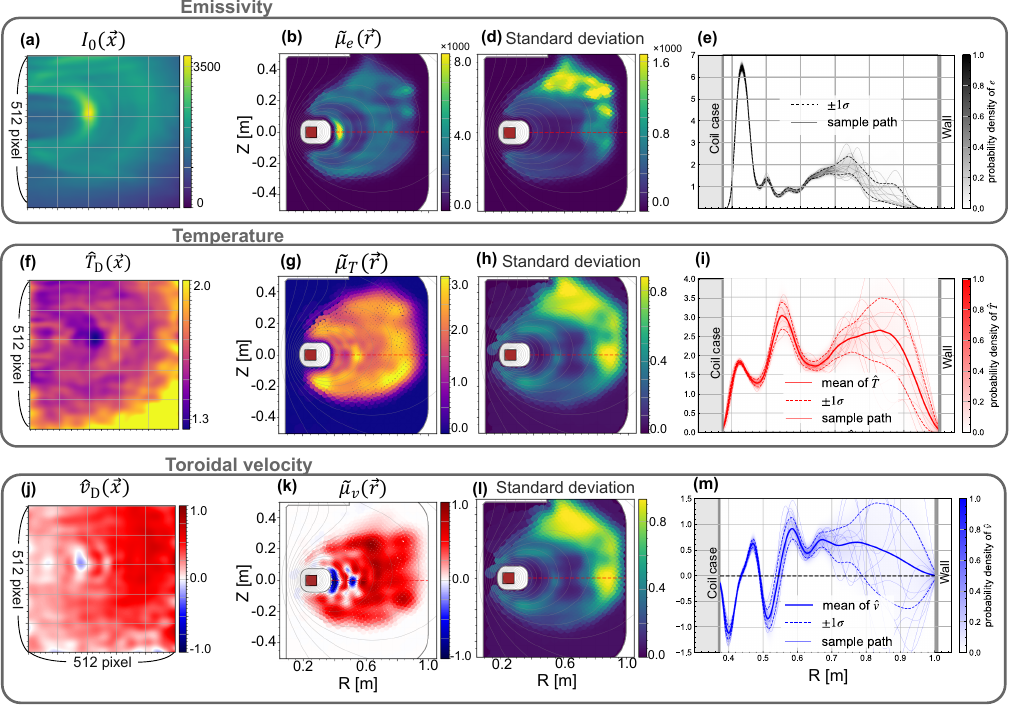}
    \caption{\label{fig: CIS_real_result_summary} Tomographic results with experimental data in RT-1. The first column, (a), (f), and (j), shows projected emissivity, ion temperature, and toroidal velocity, respectively. The second column, (b), (g), and (k), shows the mean of the posterior probabilities of emissivity, ion temperature, and toroidal velocity, respectively. The third column, (c), (h), and (l), shows the standard deviations of each physical quantity. The fourth column, (d), (i), and (m), shows the radial profiles of the mean distributions, where solid lines are the mean values, dashed lines are the \(-\sigma\) and \(+\sigma\) regions of each posterior probability, and thin colored lines are sample paths.}
\end{figure*}
In this tomography, instead of using Eq.~\eqref{eq: sigma_g}, the sigma matrix for the likelihood function, \( \Sigma_g \), is defined as the following diagonal matrix:
\begin{eqnarray}\label{eq: sigma_g_exp}
    \Sigma_g = \sigma_{g}^2 \hat{\sigma}_{g}^2(\vec{x}),
\end{eqnarray}
where the relative sigma image \( \hat{\sigma}_g(\vec{x}) \) is shown in Fig.~\ref{fig: CIS_exp_images}(c). This image assigns larger values to regions affected by complex reflections, such as the inside of a viewing port or walls with positive Gaussian curvature. Unfortunately, unlike in Fig.~\ref{fig: evidence_map}, the sigma scale could not be optimized by evidence maximization; both \( \sigma_{g} \) and \( \hat{\ell}_{F} \) converged to zero. This occurs because the diagonal model for \( \Sigma_{g} \) fits only white noise and cannot capture systematic errors such as reflections and geometric distortions. The Fourier-denoised input image therefore drives the sigma scale toward zero and causes the length scale to overfit systematic noise. Accounting for such systematic errors would require a model for how reflections and geometric distortions propagate into \( \Sigma_{g} \), which is beyond the scope of this paper. Instead, we fixed the sigma scale at \( 10\% \) and optimized only the length-scale factor.

Figure~\ref{fig: CIS_real_result_summary} shows the results of the tomography, consisting of the posterior distributions of emissivity, ion temperature, and toroidal velocity when the length scale factor is optimized to 1.97. Panels~(a), (f), and (j) show the projected emissivity, ion temperature, and toroidal velocity, respectively, obtained using Eq.~\eqref{eq: signals of CIS}. Panels~(b), (g), and (k) show the mean distributions of emissivity, ion temperature, and toroidal velocity, respectively. Panels~(c), (h), and (l) show the standard deviations of each physical quantity. The large standard deviations in the upper right regions of the poloidal cross-section are due to the relatively small number of rays or the propagation of the relative sigma image. From the radial profiles shown in panels~(d), (i), and (m), the peak of the ion temperature and the sign structure of the ion velocity are observed in the region of \( 0.4\,\mathrm{m} < R < 0.6\,\mathrm{m} \). Also, the variances of \( T_i \) and \( V_i \) tend to be larger when the emissivity is relatively small, as in the case of the phantom data. Additionally, all physical quantities have large variances in the region of \( R > 0.7\,\mathrm{m} \), which is due to the fact that the LOS do not pass tangentially through this region, as shown in Fig.~\ref{fig: top_view_CIS}.

\subsection{Discussion}\label{sec:disccusion}

\paragraph*{Model scope.} Although toroidal symmetry is assumed here for the RT-1 demonstration, the Bayesian tomographic framework itself does not rely on this assumption. The method can be extended to fully three-dimensional reconstructions by increasing the number of velocity components and state variables in the forward model.

\paragraph*{Multiplet spectra.} In general, impurity lines such as carbon (C~III) are often used for Doppler spectroscopy in the SOL in CIS diagnostics~\cite{Howard2011-ws,Silburn_MAST_2014,pub.1074243224,Perseo_Gradic_2020}. In such cases, it is necessary to consider multiplet spectra. As discussed in Appendix~\ref{app:derive_multiplet}, the effect of multiplets can be incorporated as a factor of complex numbers \( \gamma^\mathrm{mult}(\vec{x}) \) in the projection equation. As long as the intensity ratios of the spectral lines can be assumed to be constant along the LOS, the multiplet effect does not affect the integrand function of the projection equation.

\paragraph*{Zeeman effect.} The Zeeman effect should be considered in high magnetic field confinement devices for the temperature reconstruction~\cite{Gradic_2021,Gradic2022-ge,Kriete2024-hd}. It has been suggested that the Zeeman splitting can be approximated by a pseudo-temperature, such as \( \hat{T}_\mathrm{total}(\vec{r}) \simeq \hat{T}_\mathrm{Zeeman}(\vec{r}) + \hat{T}_i(\vec{r}) \), where \( \hat{T}_\mathrm{Zeeman} \) represents the contribution from the Zeeman effect. 

\paragraph{Computational cost.} The computational complexity of the present implementation is mainly determined by the Gaussian process inference and the iterative optimization required for the Laplace approximation. The computational cost scales primarily with the cubic complexity of the Gaussian process covariance inversion, which is determined by the number of inducing points. In the present RT-1 application, the number of inducing points was approximately 2000, and the reconstruction converged within a practical computational time on a standard desktop PC.

\section{Conclusion}\label{sec:conclude}
We have developed a nonlinear Bayesian tomographic framework for Doppler spectral imaging that simultaneously reconstructs emissivity, ion temperature, and flow velocity. By combining Gaussian-process priors with a nonlinear forward model and a Laplace approximation, the method enables joint inference of multiple physical quantities while preserving the underlying integral measurement equations.

The proposed approach addresses several long-standing limitations of conventional CIS tomography. In particular,
the method remains applicable in regimes with strong Doppler shifts and temperature variations, where linearized
reconstruction techniques fail. The Bayesian formulation further provides quantitative uncertainty estimates,
allowing objective assessment of reconstruction reliability.

Validation with phantom data demonstrated that accurate reconstruction is achievable even under strongly nonlinear conditions. Application to experimental measurements in the RT-1 device revealed spatial structures of ion temperature and toroidal flow characteristic of magnetospheric plasma.

Although demonstrated here for CIS, the framework is not restricted to plasma diagnostics. The methodology is directly applicable to a wide class of Doppler-based imaging systems across astrophysical observations, atmospheric remote sensing, and biomedical flow diagnostics.

The nonlinear Gaussian-process tomography presented here therefore establishes a general inverse framework for Doppler spectral imaging and provides a statistically rigorous tool for reconstructing velocity and temperature fields in complex measurement environments.

Because the framework relies only on the forward projection operator that relates local variables to observations, the method is not restricted to a specific diagnostic configuration. 
The approach can therefore be applied to different viewing geometries and measurement systems as long as the corresponding forward model is available.

The Bayesian formulation naturally incorporates measurement uncertainty through the likelihood function, while the Gaussian process prior provides spatial regularization of the reconstructed fields. As a result, the framework remains stable even in situations with limited viewing geometry or moderate measurement noise, which are common in practical spectroscopic diagnostics.

\section*{Acknowledgments}

We thank the RT-1 team for their essential support in conducting the experiments and collecting the data. This work was supported by JSPS KAKENHI Grant Nos. 19KK0073 and 23K25857.

\appendix
\appendix

\section{List of Main Quantities}\label{app:main_quantities}

This appendix summarizes the key symbols and definitions used throughout the CIS tomographic model. Tables~\ref{tab:main_quantities_cis} and \ref{tab:main_quantities_gp} list the CIS observables and the parameters associated with the Gaussian-process priors, respectively.

\begin{table}[htbp]
    \caption{\label{tab:main_quantities_cis}Summary of CIS observables, measurement noise covariances, and geometric factors for nonlinear Bayesian reconstruction.}
\begin{ruledtabular} 
    \begin{tabular}{ll}
        Symbol & Meaning \\
        \hline
        $\vec{r}$ & Position in the plasma reconstruction domain \\
        $\vec{x}$ & Pixel coordinate on the image sensor \\
        $e(\vec{r})$ & Local emissivity \\
        $T_{\mathrm{i}}(\vec{r})$ & Ion temperature \\
        $\bm{v}_{\mathrm{i}}(\vec{r})$ & Ion flow velocity \\
        $S_{\mathrm{CIS}}(\vec{x})$ & CIS output interferogram \\
        $I_0(\vec{x})$ & Observed total intensity (zero-order fringe component) \\
        $I_{\mathrm{D}}(\vec{x})$ & Complex fringe visibility (modulation component) \\
        $I_{\mathrm{Re}}^{\mathrm{obs}}, I_{\mathrm{Im}}^{\mathrm{obs}}$ & Observed quadrature components (Real/Imaginary) \\
        $\hat{T}(\vec{r})$ & Normalized ion temperature $T_{\mathrm{i}}/T_{\mathrm{c}}$ \\
        $\hat{\bm{v}}(\vec{r})$ & Normalized ion velocity $\bm{v}_{\mathrm{i}}/v_{\mathrm{c}}$ \\
        $\hat{e}(\vec{r})$ & Log-emissivity, $\hat{e} := \log e$ \\
        $\hat{a}(\vec{r})$ & Local amplitude, $\hat{a} := \hat{e} - \hat{T}$ \\
        $T_{\mathrm{c}}, v_{\mathrm{c}}$ & Characteristic temperature and velocity \\
        $\hat{N}(\vec{x})$ & Normalized group delay of the interferometer \\
        $L(\vec{x})$ & Line of sight (LOS) corresponding to pixel $\vec{x}$ \\
        $\hat{\bm{l}}$ & Unit vector along the LOS \\
        $H_{ij}$ & Geometry matrix (path length of ray $i$ through cell $j$) \\
        $\Theta_{ij}$ & Directional cosine factor for the projection $\bm{v}\cdot\hat{\bm{l}}$ \\
    \end{tabular}
    \end{ruledtabular}
\end{table}

\begin{table}[htbp]
\caption{\label{tab:main_quantities_gp}Parameters and definitions for Gaussian-process (GP) priors and tomographic inference.}
    \begin{ruledtabular}
    \begin{tabular}{ll}
        Symbol & Meaning \\
        \hline
        $\mathcal{N}(\mu,\Sigma)$ & normal distribution with mean \(\mu\) and covariance \(\Sigma\)\\
        $k(\vec{r}_i,\vec{r}_j)$ & GP kernel (covariance function) \\
        $\sigma_f$ & Signal variance (hyperparameter) \\
        $\Sigma_{\ell}(\vec{r})$ & Spatially varying length-scale matrix \\
        $\ell(\vec{r})$ & Local (position-dependent) GP length scale \\
        $\ell'(\vec{r})$ & Reference length-scale spatial profile \\
        $\hat{\ell}_F$ & Scaling factor for the length-scale profile \\
        $K_e, K_T, K_v$ & GP prior covariance matrices for $\hat{\bm{e}}$, $\hat{\bm{T}}$, and $\hat{\bm{v}}$ \\
        $\bm{\mu}_{\mathrm{pri}, e/T/v}$ & GP prior mean vectors \\
        $\Sigma_g$ & Likelihood noise covariance matrix \\
        $\sigma_g$ & Global noise scale for the CIS data \\
    \end{tabular}
    \end{ruledtabular}
\end{table}

\section{Derivation of the Projection Equation for CIS}\label{app:derive_cis}

According to previous research~\cite{Howard_2003,Howard2010-fl}, the CIS technique uses the principle of interference. The output signal \( S_\mathrm{CIS} \) is the sum of the power of the incident light and its autocorrelation at a certain delay time \( \tau \).

Given the power spectral density function of the coherent light \( \psi(\hat{\lambda}) \), where \( \hat{\lambda} \) is the normalized wavelength defined as \( \hat{\lambda} = \frac{\lambda - \lambda_0}{\lambda_0} \), the interfered signal is written as follows:
\begin{eqnarray}
 S_\mathrm{CIS}[\psi] = I_0[\psi] + \mathrm{Re}\left( \Gamma[\psi] \right),
\end{eqnarray}
where \( I_0[\psi] \) is the total energy of the spectrum, and \( \Gamma[\psi] \) is the Fourier transform of the spectrum according to the Wiener–Khinchin theorem~\cite{Bracewell2000}. The phase shift is given by
\begin{equation}\label{eq: phase_shift}
    2\pi \nu \tau = \phi_{0} - 2\pi \hat{N} \hat{\lambda},
\end{equation}
where \( \phi_{0} \) is a constant related to the interferometer, and \(\nu\) is the frequency of the light. 

The definitions of \( I_0[\psi] \) and \( \Gamma[\psi] \) are:
\begin{eqnarray}
 I_0[\psi] &=& \int_{-\infty}^{\infty} \psi(\hat{\lambda})\, \mathrm{d}\hat{\lambda},\label{eq: def_of_I0_psi}\\
 \Gamma[\psi] &=& \exp{[i\phi_0 ]} \int_{-\infty}^{\infty} \psi(\hat{\lambda}) \exp{[ -2\pi i\hat{N}\hat{\lambda}]}\, \mathrm{d}\hat{\lambda}.\label{eq: def_of_Gamma_psi}
\end{eqnarray}

Under the assumption that the spectrum of local emissivity has a single Gaussian profile~\cite{Fonck1984} with local amplitude \( e(\vec{r}) \), local wavelength shift \( \hat{\lambda}_\mathrm{D}(\vec{r}) \), and local width \( \delta_\mathrm{D}(\vec{r}) \), the spectrum of the incident light is obtained as the result of the line integral along the \( L(\vec{x}) \) and is expressed as:
\begin{eqnarray}\label{eq: integral_local_spectrum}
  \psi(\vec{x},\hat{\lambda}) = \int_{L(\vec{x})} \frac{e(\vec{r})}{\sqrt{2\pi \delta^2_\mathrm{D}(\vec{r})}} \exp\left[ -\frac{ ( \hat{\lambda} - \hat{\lambda}_\mathrm{D}(\vec{r}) )^2 }{ 2\delta_\mathrm{D}^2(\vec{r}) } \right] \mathrm{d}l.
\end{eqnarray}

Substituting Eq.~\eqref{eq: integral_local_spectrum} into Eqs.~\eqref{eq: def_of_I0_psi} and~\eqref{eq: def_of_Gamma_psi}, \( I_0(\vec{x}) \) and \( \Gamma[\psi_{\vec{x}}] \) are derived by interchanging the order of integration:
\begin{eqnarray}
 I_0[\psi_{(\vec{x})}] \equiv I_0(\vec{x}) = 
  \int_{L(\vec{x})} e(\vec{r}) \, \mathrm{d}l, 
\end{eqnarray}
and
\begin{eqnarray}
\Gamma[\psi_{(\vec{x})}]
 &=& \exp[ i\phi_0(\vec{x}) ] \int_{L(\vec{x})} \int_{-\infty}^{\infty} \frac{e(\vec{r})}{\sqrt{2\pi \delta^2_\mathrm{D}(\vec{r})}}  \exp\left[ -\frac{ ( \hat{\lambda} - \hat{\lambda}_\mathrm{D}(\vec{r}) )^2 }{ 2\delta_\mathrm{D}^2(\vec{r})} \right]  \nonumber \\&\times& \exp[ -2\pi i\hat{N} (\vec{x})\hat{\lambda}  ] \mathrm{d}\hat{\lambda}  \mathrm{d}l \nonumber\\
 &=& \exp[ i\phi_0(\vec{x}) ]  \int_{L(\vec{x})} e(\vec{r}) \nonumber\\
 &\times& \exp\left[ -\frac{1}{2} (2\pi \hat{N}(\vec{x}))^2 \delta_\mathrm{D}^2(\vec{r}) + i 2\pi \hat{N}(\vec{x}) \hat{\lambda}_\mathrm{D}(\vec{r}) \right] \mathrm{d}l \nonumber\\
 &\equiv& \exp[ i\phi_0(\vec{x}) ]  I_\mathrm{D}(\vec{x}). \label{eq: transform_ID}
\end{eqnarray}

Under the assumption of a Maxwellian distribution~\cite{Hutchinson2002}, the local Doppler shift and broadening are related to the ion velocity \( \bm{v}_\mathrm{i}(\vec{r}) \) and ion temperature \( T_\mathrm{i}(\vec{r}) \) by:
\begin{eqnarray}\label{eq: to_T_and_V}
    2\pi \hat{N} \hat{\lambda}_\mathrm{D}(\vec{r}) = \frac{ \bm{v}_\mathrm{i}(\vec{r}) \cdot \hat{\bm{l}} }{ v_\mathrm{c} }, \quad
    \frac{1}{2} (2\pi \hat{N})^2 \delta^2_\mathrm{D}(\vec{r}) = \frac{ T_\mathrm{i}(\vec{r}) }{ T_\mathrm{c} },
\end{eqnarray}
where \( v_\mathrm{c} \) and \( T_\mathrm{c} \) are characteristic velocity and temperature, respectively, and \( \hat{\bm{l}} \) is the unit vector along the LOS. Substituting Eq.~\eqref{eq: to_T_and_V} into Eq.~\eqref{eq: transform_ID}, we obtain the projection equation used in CIS tomography.

\section{Projection Equation for the Multiplet Spectrum}\label{app:derive_multiplet}

We consider the function of the multiplet spectrum without Doppler shift and broadening as~\cite{Gradic_2021}:
\begin{eqnarray}
    \psi^\mathrm{multi}(\lambda) = e \sum_{k} C_k \delta(\lambda - \lambda_k),
\end{eqnarray}
where \( e \) is the total emissivity, \( \lambda_k \) is the wavelength of each line, and \( C_k \) is the relative intensity of each line satisfying \( \sum_k C_k = 1 \).

Under the assumption that \( C_k \) and \( \lambda_k \) are known and constant at each position \( \vec{r} \), the shifted and broadened spectrum \( \psi^\mathrm{multi}(\vec{r}, \hat{\lambda}) \) at each \( \vec{r} \) is formulated as:
\begin{eqnarray}\label{eq: multi_local_spectrum}
    \psi^\mathrm{multi}(\vec{r}, \hat{\lambda}) = e(\vec{r}) \sum_{k} \frac{C_k}{\sqrt{2\pi \delta^2_\mathrm{D}(\vec{r})}} \exp\left( -\frac{ \left( \hat{\lambda} - \hat{\lambda}_k - \hat{\lambda}_\mathrm{D}(\vec{r}) \right)^2 }{ 2\delta_\mathrm{D}^2(\vec{r}) } \right),\nonumber\\
\end{eqnarray}
where \( \hat{\lambda}_k = \frac{ \lambda_k - \lambda_0 }{ \lambda_0 } \).
Substituting \( \psi^\mathrm{multi}(\hat{\lambda}) = \int \psi^\mathrm{multi}(\vec{r}, \hat{\lambda}) \, \mathrm{d}l \) into Eq.~\eqref{eq: def_of_Gamma_psi}, we obtain:
\begin{eqnarray}
    \Gamma[\psi^\mathrm{multi}_{(\vec{x})}] &=& \exp\left[ i\phi_0(\vec{x}) \right] \gamma^\mathrm{multi}(\vec{x}) \nonumber\\
    &\times& \int_{L(\vec{x})} e(\vec{r}) \exp\left( -\frac{ T_\mathrm{i}(\vec{r}) }{ T_\mathrm{c} } + i \frac{ \bm{v}_\mathrm{i}(\vec{r}) \cdot \hat{\bm{l}} }{ v_\mathrm{c} } \right) \mathrm{d}l, \nonumber\\
    \text{where} \quad \gamma^\mathrm{multi}(\vec{x}) &=& \sum_k C_k \exp\left[ i 2\pi \hat{N}(\vec{x}) \hat{\lambda}_k \right].\label{eq: gamma_multiplet_appendix}
\end{eqnarray}
Thus, in this case, the influence of the multiplet is summarized in the function \( \gamma^\mathrm{multi}(\vec{x}) \) and does not affect the integral process.

\section{Calculus for the log-posterior function of CIS}\label{app:derive_Phi_cis}

Using the Einstein summation convention, the gradient of the log-posterior distribution \( \nabla \Psi^\mathrm{CIS}(\tilde{\bm{a}}, \tilde{\bm{v}}) \) is derived for each \( \tilde{a}_j \) and \( \tilde{v}_j \) as follows:
\begin{eqnarray}
  \frac{\partial}{\partial \tilde{a}_j} \Psi^\mathrm{CIS}(\tilde{\bm{a}}, \tilde{\bm{v}})
    &=& - \epsilon^\mathrm{C}_i \sigma^{-2}_{i} A^\mathrm{C}_{ij}
        - \epsilon^\mathrm{S}_i \sigma^{-2}_{i} A^\mathrm{S}_{ij}\nonumber\\
       && - \left[ \Sigma^{-1}_{a} (\tilde{\bm{a}} - \bm{\mu}^\mathrm{pri}_{a}) \right]_j, \label{eq: dPsida_appendix}\\
    \frac{\partial}{\partial \tilde{v}_j} \Psi^\mathrm{CIS}(\tilde{\bm{a}}, \tilde{\bm{v}})
    &=& + \epsilon^\mathrm{C}_i \sigma^{-2}_{i} \Theta_{ij} A^\mathrm{S}_{ij}
        - \epsilon^\mathrm{S}_i \sigma^{-2}_{i} \Theta_{ij} A^\mathrm{C}_{ij}\nonumber\\
       && - \left[ K^{-1}_{v} (\tilde{\bm{v}} - \bm{\mu}^\mathrm{pri}_{v}) \right]_j, \label{eq: dPsidv_appendix}
\end{eqnarray}
where the summation over \( i \) is implied, and the covariance matrix \( \Sigma_g \) is assumed to be diagonal, \( [\Sigma_g]_{ij} = \sigma^2_i \delta_{ij} \). The variables \( \epsilon^\mathrm{C}, \epsilon^\mathrm{S}, A^\mathrm{C} \), and \( A^\mathrm{S} \) are given by:
\begin{eqnarray*}
    A^\mathrm{C}_{ij} &:=& H_{ij} \exp\left( \tilde{a}_{j} \right) \cos\left( \Theta_{ij} \tilde{v}_{j} \right), \\
    A^\mathrm{S}_{ij} &:=& H_{ij} \exp\left( \tilde{a}_{j} \right) \sin\left( \Theta_{ij} \tilde{v}_{j} \right), \\
    \epsilon^\mathrm{C}_i &:=& g_{\mathrm{C}, i}(\tilde{\bm{a}}, \tilde{\bm{v}}) - I_{\mathrm{Re}, i}^\mathrm{obs}, \\
    \epsilon^\mathrm{S}_i &:=& g_{\mathrm{S}, i}(\tilde{\bm{a}}, \tilde{\bm{v}}) - I_{\mathrm{Im}, i}^\mathrm{obs}.
\end{eqnarray*}
As a result, the following equations hold:
\begin{eqnarray*}
    \frac{\partial g_{\mathrm{C}, j}(\tilde{\bm{a}}, \tilde{\bm{v}})}{\partial \tilde{a}_i}
    &=& + A^\mathrm{C}_{ji}, \quad
    \frac{\partial g_{\mathrm{S}, j}(\tilde{\bm{a}}, \tilde{\bm{v}})}{\partial \tilde{a}_i}
    = + A^\mathrm{S}_{ji}, \\
    \frac{\partial g_{\mathrm{C}, j}(\tilde{\bm{a}}, \tilde{\bm{v}})}{\partial \tilde{v}_i}
    &=& - \Theta_{ji} A^\mathrm{S}_{ji}, \quad
    \frac{\partial g_{\mathrm{S}, j}(\tilde{\bm{a}}, \tilde{\bm{v}})}{\partial \tilde{v}_i}
    = + \Theta_{ji} A^\mathrm{C}_{ji}, \\
    \frac{\partial A^\mathrm{C}_{kj}}{\partial \tilde{a}_i}
    &=& + A^\mathrm{C}_{kj} \delta_{ij}, \quad
    \frac{\partial A^\mathrm{S}_{kj}}{\partial \tilde{a}_i}
    = + A^\mathrm{S}_{kj} \delta_{ij}, \\
    \frac{\partial A^\mathrm{C}_{kj}}{\partial \tilde{v}_i}
    &=& - \Theta_{kj} A^\mathrm{S}_{kj} \delta_{ij}, \quad
    \frac{\partial A^\mathrm{S}_{kj}}{\partial \tilde{v}_i}
    = + \Theta_{kj} A^\mathrm{C}_{kj} \delta_{ij}.
\end{eqnarray*}
Using the above equations, the Hessian of the log-posterior distribution \( \nabla^2 \Psi^\mathrm{CIS}(\tilde{\bm{a}}, \tilde{\bm{v}}) \) is derived for each \( i, j \) as follows:
\begin{eqnarray}
   \frac{\partial^2}{\partial \tilde{a}_i \partial \tilde{a}_j} \Psi^\mathrm{CIS}(\tilde{\bm{a}}, \tilde{\bm{v}}) &=&
   - A^\mathrm{C}_{ki} \sigma^{-2}_{k} A^\mathrm{C}_{kj}
   - A^\mathrm{S}_{ki} \sigma^{-2}_{k} A^\mathrm{S}_{kj} \nonumber\\
   && - \delta_{ij} \epsilon^\mathrm{C}_k \sigma^{-2}_{k} A^\mathrm{C}_{kj}
      - \delta_{ij} \epsilon^\mathrm{S}_k \sigma^{-2}_{k} A^\mathrm{S}_{kj} \nonumber\\
   && - [\Sigma^{-1}_{a}]_{ij}, \label{eq: ddPsidada_appendix}\\
   \frac{\partial^2}{\partial \tilde{v}_i \partial \tilde{v}_j} \Psi^\mathrm{CIS}(\tilde{\bm{a}}, \tilde{\bm{v}}) &=&
   - \Theta_{ki} A^\mathrm{C}_{ki} \sigma^{-2}_{k} \Theta_{kj} A^\mathrm{C}_{kj}
   - \Theta_{ki} A^\mathrm{S}_{ki} \sigma^{-2}_{k} \Theta_{kj} A^\mathrm{S}_{kj} \nonumber\\
   && + \delta_{ij} \epsilon^\mathrm{C}_k \sigma^{-2}_{k} \Theta_{kj}^2 A^\mathrm{C}_{kj}
      + \delta_{ij} \epsilon^\mathrm{S}_k \sigma^{-2}_{k} \Theta_{kj}^2 A^\mathrm{S}_{kj} \nonumber\\
   && - [K^{-1}_{v}]_{ij}, \label{eq: ddPsidvdv_appendix}\\
   \frac{\partial^2}{\partial \tilde{a}_i \partial \tilde{v}_j} \Psi^\mathrm{CIS}(\tilde{\bm{a}}, \tilde{\bm{v}}) &=&
   + A^\mathrm{C}_{ki} \sigma^{-2}_{k} \Theta_{kj} A^\mathrm{S}_{kj}
   - A^\mathrm{S}_{ki} \sigma^{-2}_{k} \Theta_{kj} A^\mathrm{C}_{kj} \nonumber\\
   && + \delta_{ij} \epsilon^\mathrm{C}_k \sigma^{-2}_{k} \Theta_{kj} A^\mathrm{S}_{kj}
      - \delta_{ij} \epsilon^\mathrm{S}_k \sigma^{-2}_{k} \Theta_{kj} A^\mathrm{C}_{kj} \nonumber\\
   &=&  \frac{\partial^2}{\partial \tilde{v}_j \partial \tilde{a}_i} \Psi^\mathrm{CIS}(\tilde{\bm{a}}, \tilde{\bm{v}}), \label{eq: ddPsidadv_appendix}
\end{eqnarray}
where the summation over \( k \) is implied.

\bibliographystyle{unsrt}  
\bibliography{your-bib-file}  
\end{document}